\documentclass[modern]{aastex631}

\usepackage{chemformula}
\usepackage{tabularx}
\usepackage{svg}
\usepackage{csvsimple}
\usepackage{graphicx}

\begin{document}

\title{Inner Edge Habitable Zone Limits Around Main Sequence Stars: Cloudy Estimates}

\correspondingauthor{James D. Windsor}
\email{jdw472@nau.edu}

%
\author[0000-0001-8522-3788]{James D. Windsor}
\affiliation{Department of Astronomy and Planetary Science, Northern Arizona University, Flagstaff, AZ 86011, USA}
\affiliation{Habitability, Atmospheres, and Biosignatures Laboratory, University of Arizona, Tucson, AZ 85721, USA}
\affiliation{NASA Nexus for Exoplanet System Science, Virtual Planetary Laboratory Team, Box 351580, University of Washington, Seattle, Washington 98195, USA}

\author[0000-0002-3196-414X]{Tyler D. Robinson}
\affiliation{Lunar \& Planetary Laboratory, University of Arizona, Tucson, AZ 85721 USA}
\affiliation{Department of Astronomy and Planetary Science, Northern Arizona University, Flagstaff, AZ 86011, USA}
\affiliation{Habitability, Atmospheres, and Biosignatures Laboratory, University of Arizona, Tucson, AZ 85721, USA}
\affiliation{NASA Nexus for Exoplanet System Science, Virtual Planetary Laboratory Team, Box 351580, University of Washington, Seattle, Washington 98195, USA}

\author[0000-0002-5893-2471]{Ravi kumar Kopparapu}
\affiliation{NASA Goddard Space Flight Center 8800 Greenbelt Road Greenbelt, MD 20771, USA}

\author[0000-0001-8106-6164]{Arnaud Salvador}
\affiliation{Lunar \& Planetary Laboratory, University of Arizona, Tucson, AZ 85721 USA}
\affiliation{Habitability, Atmospheres, and Biosignatures Laboratory, University of Arizona, Tucson, AZ 85721, USA}
\affiliation{NASA Nexus for Exoplanet System Science, Virtual Planetary Laboratory Team, Box 351580, University of Washington, Seattle, Washington 98195, USA}

\author[0000-0003-3099-1506]{Amber V. Young}
\affiliation{NASA Goddard Space Flight Center 8800 Greenbelt Road Greenbelt, MD 20771, USA}
\affiliation{Department of Astronomy and Planetary Science, Northern Arizona University, Flagstaff, AZ 86011, USA}
\affiliation{Habitability, Atmospheres, and Biosignatures Laboratory, University of Arizona, Tucson, AZ 85721, USA}
\affiliation{NASA Nexus for Exoplanet System Science, Virtual Planetary Laboratory Team, Box 351580, University of Washington, Seattle, Washington 98195, USA}

\author[0000-0002-1386-1710]{Victoria S. Meadows}
\affiliation{Department of Astronomy and Astrobiology Program, University of Washington, Box 351580, Seattle, Washington 98195, USA}
\affiliation{NASA Nexus for Exoplanet System Science, Virtual Planetary Laboratory Team, Box 351580, University of Washington, Seattle, Washington 98195, USA}



\begin{abstract}

Understanding the limits of rocky planet habitability is one of the key goals of current and future exoplanet characterization efforts. An intrinsic concept of rocky planet habitability is the Habitable Zone. To date, the most widely used estimates of the Habitable Zone are based on cloud-free, one-dimensional (vertical) radiative-convective climate model calculations. However, recent three-dimensional global climate modeling efforts have revealed that rocky planet habitability is strongly impacted by radiative cloud feedbacks, where computational expense and model limitations can prevent these tools from exploring the limits of habitability across the full range of parameter space. We leverage a patchy cloud one-dimensional radiative-convective climate model with parameterized cloud microphysics to investigate Inner Edge limits to the Habitable Zone for main sequence stars ($T_{\rm eff}$ =  2600--7200\,K). We find that Inner Edge limits to the Habitable Zone can be 3.3 and 4.7 times closer than previous cloud-free estimates for Earth- and super-Earth-sized worlds, respectively, depending on bulk cloud parameters (e.g., fractional cloudiness and sedimentation efficiency). These warm, moist Inner Edge climates are expected to have extensive cloud decks that could mute deep atmosphere spectral features. To aid in rocky planet characterization studies, we identify the potential of using \ch{CO2} absorption features in transmission spectroscopy as a means of quantifying cloud deck height and cloud sedimentation efficiency. Moist greenhouse climates may represent key yet poorly understood states of habitable planets for which continued study will uncover new insights into the search and characterization of habitable worlds.


\end{abstract}



\section{Introduction} \label{sec:intro}


Based on the assumption that the origin and maintenance of life requires a stable source of surface liquid water, it has been postulated that life, and therefore habitability, is more likely to arise on terrestrial planets within the Habitable Zone. Planet composition studies reveal that worlds with planetary radii smaller than roughly 1.6\,R$_{\rm \oplus}$ are unlikely to be gas-dominated \citep{rogers2015} and planet occurrence studies propose that rocky worlds are some of the more common type of planet arising in our galaxy \citep{brysonetal2021}. Further, a number of early studies have identified rocky surfaces as a fundamental planetary characteristic for the emergence of life \citep{dole1964,shklovsky1966}. \citet{dole1964} and \citet{shklovsky1966} anthropomorphized planetary habitability and coined the term \textit{ecosphere} to describe the planetary requirements necessary for human life. Using simple estimation methods, \citet{dole1964} predicted that the ecosphere is bounded by inner (0.725 AU) and outer (1.24 AU) orbital distance limits for sun-like stars. A more general \textit{Habitable Zone}\,---\,described first by \citet{huang1959,huang1960} and later by \citet{kasting1988,whitmire1991,kastingetal1993}\,---\,assumes that the origin an maintenance of life on planets elsewhere in our galaxy requires stable liquid water on the planetary surface. Many related studies further developed these requirements in terms of planetary surface conditions and climate characteristics \citep{kasting1988,kastingetal1993,pierrehumbert1995,renn1997,selsisetal2007,zsometal2012,kopparapu2013,yang2013,goldblattetal2013,leconteetal2013,kopparapuetal2014,wolfeandtoon2014,wolfeandtoon2015,kopparapuetal2016,kopparapuetal2017,fouchezetal2021,wayetal2022}. Recently, increasing importance has also been given to initial surface ocean formation and its relationship to young planet habitability and eventual climate evolution \citep[e.g.,][]{Hamano2013, Turbet2021, Salvador2023_SSR}. In general, the Habitable Zone can be quantified in terms of an \textit{Outer Edge} and an \textit{Inner Edge}, far and close from the host star, respectively. Each edge case is modeled for specific climatic scenarios to either maximize (Outer Edge) or minimize (Inner Edge) the atmospheric greenhouse effect on the planetary climate. Specifically, the Outer Edge maximizes greenhouse warming through a dense \ch{CO2} atmosphere \citep[at least in the classical treatment; although see][]{pierrehumbertandgaidos2011}, whereas the Inner Edge cases classically rely on water vapor radiative feedback mechanisms in steam atmospheres that culminate in an eventual ocean's worth of water loss over rapid (runaway greenhouse) or geologic timescales (moist greenhouse). 

Planets within the Habitable Zone are key targets for current and future rocky exoplanet characterization efforts \citep{wordsworthetal2022}, and Inner Edge planets are likely more observationally accessible. However, current observational capabilities are strongly sensitive to planetary orbital distance from the host star \citep{kopparapuetal2017}. Within the Habitable Zone, characterizations of high molecular weight secondary atmospheres with the NASA's \textit{James Webb Space Telescope} (\textit{JWST}) may require 10--100s of transits \citep{lustig-yaegeretal2019}. Within these limits, Inner Edge planets are prime targets due to their shorter orbital periods which lead to a higher potential cadence of follow up observations. However, extrasolar Inner Edge planets may have climate states rather distinct from either Earth or Venus. Fully understanding Inner Edge planets will require leveraging observation-based planetary characterization studies and theoretical numerical modeling efforts such as those employed in \citet{kasting&ackerman1986,kasting1988,kastingetal1993,pierrehumbert1995,renn1997,selsisetal2007,kitzmannetal2010,zsometal2012,wolfeandtoon2014,kopparapuetal2013,yang2013,kopparapuetal2014,kopparapuetal2016,kopparapuetal2017}. Thus, continued modeling studies of Inner Edge worlds can help to better understand the potential diversity of habitable environments in the galaxy and aid with observation planning and interpretation. Previous modeling studies suggest Inner Edge limit worlds with adequate surface water inventories may be in either a moist or runaway greenhouse state \citep{kasting&ackerman1986,kasting1988,kastingetal1993,pierrehumbert1995,renn1997,selsisetal2007,kitzmannetal2010,zsometal2012,wolfeandtoon2014,kopparapuetal2013,yang2013,kopparapuetal2014,kopparapuetal2016,kopparapuetal2017}.

\subsection{Moist Greenhouse Limit}

The \textit{moist greenhouse limit} occurs when the stratosphere of an ocean-bearing world becomes enriched with water vapor and the planet undergoes rapid photolytic hydrogen diffusion-driven water loss. The water loss transition occurs when the water vapor mixing ratio exceeds 10--20\% in the troposphere \citep{1ingersoll969,kastingetal1993}, where the cold-trapping of water vapor at the tropopause then becomes ineffective. Photolytic hydrogen diffusion water loss becomes non-negligible when the stratospheric water volume mixing ratio exceeds roughly $10^{-3}$ for an Earth-sized planet, above which Earth's entire inventory of water is lost to space within the present age of the Earth (4.6\,Gyr) or less, subsequently terminating habitability. The first Inner Edge limit calculations by \citep{kastingetal1993} standardized this threshold in terms of incident flux on a planet through leveraging a one-dimensional radiative-convective equilibrium climate model. These initial calculations reported a threshold orbital distance of 0.95 AU for an Earth-like planet around a Sun-like star and correspond to an incident flux that is 10\% higher than the current flux the Earth receives. More recent moist greenhouse Inner Edge calculations by \citet{kopparapuetal2013} updated \ch{H2O} absorption coefficient data and reported an incident flux threshold of 1.014, corresponding to an orbital distance of 0.99 AU for an Earth-like planet orbiting a Sun-like star. Finally, \citet{kopparapuetal2014} expanded planet size parameters to include rocky super-Earth worlds with planetary radii up to 1.5R$_{\rm \oplus}$ while adopting the same stratospheric saturation threshold of $10^{-3}$ for water vapor. Despite the loss of long-term habitability, worlds at the Inner Edge of the Habitable Zone likely retain stable conditions for surface water for long periods of time; thus, the moist greenhouse limit is a conservative estimate for habitability limits.

\subsection{Runaway Greenhouse Limit}

A planet that surpasses the moist greenhouse threshold will reach a climate state in which the strong mid-infrared opacity of water vapor decouples the deep atmosphere (and surface) cooling from shortwave solar heating. Planetary energy balance can only be achieved once the surface has warmed to 1,400\,K where emission through near-infrared windows in the water vapor opacity can stabilize the thermal conditions in the deep atmosphere \citep[e.g.,][]{Goldblatt2012}. 
At planetary surface temperatures approaching, then exceeding, those of Venus, a rocky world succumbing to a global Runaway Greenhouse state is not habitable, and represents the extreme Inner Edge scenario. \citet{kastingetal1993} first reported the Runaway Greenhouse limit at an insolation of 1.41 times the current flux Earth receives. More recently, \citet{kopparapuetal2013} reported a Runaway Greenhouse incident flux threshold of 1.107 times the current flux Earth receives for Earth-sized worlds, while this limit is increased to 1.188 for super-Earths \citep{kopparapuetal2014}. 

Studies that leveraged the capabilities of three-dimensional global climate models \citep{yang2013,wolfeandtoon2014,wolfeandtoon2015,kopparapuetal2016,kopparapuetal2017} assessed planets in climate states approaching the Runaway Greenhouse Limit, and determined that climate feedbacks\,---\,previously unable to be modeled with one-dimensional climate models\,---\,may prevent the onset of a Runaway Greenhouse climate state. Primarily, water cloud driven radiative feedbacks may insulate moist greenhouse Inner Edge planets from the Runaway Greenhouse state. However, \citet{wolfeandtoon2014,wolfeandtoon2015} found that as the incident flux is increased, the planet continues to warm and eventually water clouds disappear entirely due an exponentially increasing temperature-dependant saturation vapor pressure threshold. After cloud loss, the temperature-dependent photolytic hydrogen diffusion-driven water loss mechanism is expected to rapidly dissipates water in the planetary atmosphere before the establishment of a Runaway Greenhouse state.

\subsection{The Importance of Using Realistic Clouds in Models of Inner Edge Limit Worlds}

The majority of current Inner Edge limit calculations approximate cloud effects via enhanced surface albedo, and neglect other cloud radiative impacts on planetary climate, potentially leading to significant differences in our understanding of the location of the Inner Edge limit. 
One key climatic effect of clouds on the location(s) of the Habitable Zone limits is the radiative cooling driven by the cloud albedo effect \citep{lacis2012}. In earlier one-dimensional studies \citep{kasting1988,kastingetal1993,kopparapuetal2013,kopparapuetal2014}, a fixed cloud albedo effect is accounted for by tuning a gray surface albedo to reproduce the Modern Earth's surface temperature of 288.5\,K when an Earth-like atmospheric composition is adopted. This approach neglects important radiative feedback mechanisms from clouds\,---\,the cloud greenhouse effect (longwave) \citep{lacis2012,windsoretal2022}, and the cloud albedo effect (shortwave) \citep{lacis2012,windsoretal2022}. In tandem, these cloud feedbacks are expected to push the location(s) of the Inner Edge limits closer to their host stars \citep{kastingetal1993,pierrehumbert1995,renn1997,selsisetal2007,kitzmannetal2010,kopparapuetal2013,yang2013,kopparapuetal2014,wolfeandtoon2014,wolfeandtoon2015,kopparapuetal2016,kopparapuetal2017}. Cloud radiative effects, including absorption and scattering, are intrinsic to physical cloud properties such as cloud composition, horizontal spatial distribution, vertical extent, droplet number density, droplet size distribution, and sedimentation dynamics. Thus, incorporating more physically rigorous cloud models into broadly-applicable one-dimensional planetary climate models can help to greatly increase our understanding of Inner Edge climate states across a diversity of worlds. 

\subsection{Previous Cloudy Inner Edge Habitable Zones}

Fully resolved cloud models capable of self-consistently modeling the cloud radiative effects aforementioned are generally too computationally expensive  to leverage for large-scale, gridded, modeling approaches \citep{windsoretal2022}. Several previous cloudy Habitable Zone studies have attempted to model the climatic influence of clouds arising in planetary atmospheres via simplified cloud treatments \citep{kasting1988,pierrehumbert1995,selsisetal2007,kitzmannetal2010,zsometal2012}. However, simplified cloud modeling approaches may under (or over) estimate cloud-climate effects \citep{zsometal2012,gaoetal2021,windsoretal2022}. 

For instance, \citet{kasting1988} adopted a homogeneous cloud structure with spherical water droplets of a mean radius of 5\,{\textmu}m in a lognormal size distribution. The thickness of the cloud layer was taken to be approximately one pressure scale height, the cloud liquid water content was weighted by the atmospheric layer density, and cloud altitudes were treated as a free parameter. Here, \citet{kasting1988} reported that the inclusion of clouds in a warm ($T_{\rm 0}$ = 647~K, $P_{\rm 0}$=203~bar), close-to-runaway modeling case increased the required Modern Earth scaled incident flux ($S_{\rm eff}$) to induce a runaway greenhouse from 1.4\,$S_{\rm eff}$ (in the cloud-free case) to 2.2\,$S_{\rm eff}$ in the 50\% clouded case and 4.8 $S_{\rm eff}$ in the 100\% clouded case. These early findings from \citet{kasting1988} suggested that appropriate cloud treatments should be fruitful in expanding the Habitable Zone closer to the host star. 

More recent studies explored the role of clouds on the Inner Edge of the Habitable Zone \citep{selsisetal2007,kitzmannetal2010,yang2013,wolfeandtoon2014,wolfeandtoon2015,kopparapuetal2016,kopparapuetal2017}, but, in some cases, extrapolated from limited model runs or parameterized cloud properties based on Earth. In such cases, cloud behaviors may not be applicable to novel climate states like the global moist greenhouse.  For instance, \citet{selsisetal2007} calculated the radiative influence of parameterized clouds in the shortwave regime only and argued that, in the Runaway Greenhouse limit, clouds do not contribute significantly to greenhouse warming. Here, the opacity of the atmosphere to thermal radiation due to water vapor was assumed to render any cloud greenhouse warming negligible. Nonetheless, \citet{selsisetal2007} found that infrared inactive clouds push the Inner Edge of the Habitable Zone for a Sunlike star to 0.46 AU for the 100\% clouded case, and 0.68 AU for the 50\% clouded case. \citet{kitzmannetal2010} briefly explored the influence of clouds on the Inner Edge of the Habitable Zone and found that, for their 100\% cloudy cases, the influence of clouds moved the Inner Edge of the Habitable Zone roughly 35\% closer to the host star than the cloud-free results of \citep{kopparapuetal2013}. Other studies explored the influence of clouds on the Inner Edge of the Habitable Zone for rotationally-locked worlds with three-dimensional global climate models \citep{yang2013,kopparapuetal2016,kopparapuetal2017}. They found that clouds produce a stable radiative feedback mechanism that effectively cools the planet and significantly shifts the Inner Edge limit of the Habitable Zone for stellar hosts with effective temperatures of 4,500\,K from 0.9 times the incident flux incident on Modern Earth to a range spanning 2.2--1.7 times the incident flux on Modern Earth \citep{kopparapuetal2013}.

Many such three-dimensional climate models approximate cloud radiative effects (and intrinsic microphysical properties) via adopting uniform cloud droplet sizes \citep{kopparapuetal2016,kopparapuetal2017} which results in a more than 50\% error in cloud radiative effects \citep{gaoetal2021}. Further, \citet{yang2013} tested only $P_{\rm 0}$=1~bar \ch{N2}, \ch{H2O} atmospheres, whereas rocky worlds in a close-to-runaway state are expected to have planetary surface pressures close to 200~bar, where the majority of their content is \ch{H2O} \citep{kasting1988,kastingetal1993,kopparapu2013}. There is a need to adopt a climate model with appropriate treatments of clouds, while also maintaining flexibility in atmospheric composition, planet characteristics, and computational efficiency, to re-compute the influence of clouds on the Inner Edge of the Habitable Zone. Thus, gridded modeling approaches, such as Inner Edge limit climate explorations need an intermediate complexity cloud model that self-consistently computes approximated cloud microphysics while maintaining computational efficiency \citep[e.g.,][]{windsoretal2022}.

\subsection{Summary}
In summary, the radiative influence of water clouds arising in Earth-like atmospheres is likely to significantly alter the location of the Inner Edge of the Habitable Zone. Further, the effect of clouds on the location of the Inner Edge of the Habitable Zone depends heavily on interconnected physical cloud properties, such as cloud top height, fractional cloudiness, cloud droplet size distribution, saturation physics, and sedimentation physics. Capturing each effect with fully-resolved cloud microphysics tools is computationally prohibitive, yet neglecting these cloud dynamics precludes important cloud-driven climate feedback mechanisms that arise in the unique diversity of planetary environments and stellar hosts that comprise the Inner Edge limit of terrestrial planet habitability. Here, we invoke the methods of \citet{ackemanandmarley2001,marleyetal2010} and adopt a computationally efficient parameterized cloud microphysics tool for use in one-dimensional radiative-convective climate models for rocky planets \citep{windsoretal2022}. In Section~\ref{sec:Methods} we review and formulate new methods for calculating the Inner Edge of the Habitable Zone, and include a short description of our  modeling capabilities. Section~\ref{sec:Results} showcases key findings that include cloud-driven radiative effects on the location of the Inner Edge of the Habitable Zone. Section~\ref{sec:Discussion} interprets new estimates for the Inner Edge of the Habitable Zone and frames how these compare to previous findings. Finally, Section~\ref{sec:Conclusion} summarizes the key findings of this manuscript and presents key take-away points.

\section{Methods}\label{sec:Methods}

We use the one-dimensional radiative-convective climate model introduced and validated in \citet{windsoretal2022} along with methods similar to \citet{kopparapu2013,kopparapuetal2014} to explore the moist greenhouse Inner Edge of the Habitable Zone for main sequence stellar hosts with the influence of convective water clouds with 1~bar (Earth-sized, $R_{\rm \oplus}$ =1.0, $M_{\rm \oplus}$ = 1.0), and 2.5~bar (super-Earth-sized, $R_{\rm \oplus}$ =1.5, $M_{\rm \oplus}$ = 3.5) \ch{N2} background atmospheres, and Modern Earth equivalent \ch{CO2} inventories. The radiative transfer scheme in the climate model is based on the delta-Eddington two stream solution outlined in \citet{toonetal1989}. Fluxes in the shortwave are binned into 38 equally spaced spectral bins ranging from (0.2--4.35\,{\textmu}m). Longwave fluxes are binned into 55 spectral intervals from 0--15,000\,cm$^{-1}$ and are spaced to capture key \ch{H2O} and \ch{CO2} absorption bands \citep{kasting&ackerman1986}. Absorption by \ch{CO2} and \ch{H2O} is parameterized via correlated-$k$ coefficients \citep{kopparapu2013,kopparapuetal2014,windsoretal2022} that have been updated with new linelist data from \citet{HITRAN2020} and \citet{HITEMP2020}.

As pointed out in earlier works \citep{kasting1988,kastingetal1993,selsisetal2007,kopparapu2013,yang2013,kopparapuetal2014,kopparapuetal2016,kopparapuetal2017}, the radiative influence of clouds on the Inner Edge of the Habitable Zone is considerable and difficult to capture. Previous efforts in the realm of giant planet climate modeling \citep{marleyetal2010,fortneyetal2011} and rocky planet climate modeling \citep{kitzmannetal2010,fouchezetal18,windsoretal2022} have demonstrated considerable benefit to adopting a one-dimensional patchy cloud treatment, in which the radiative effects of patchy clouds can be approximated by implementing a radiative transfer subcolumn treatment where cloudy and clear-sky radiative fluxes are fractionally weighted. In this scheme the global cloud fraction is parameterized through adopting a fractional cloudiness parameter, $f_{\rm cld}$ \citep{windsoretal2022}.

Here, rather than being prescribed, cloud optical properties are self-consistently computed via a parameterized cloud microphysics model \citep{ackemanandmarley2001,windsoretal2022}. This parameterized microphysics scheme balances the upward transport of water vapor via convective mixing with the downward transport of water vapor via condensation and sedimentation. Here, the cloud droplet size distributions are presumed to be lognormal with a free parameter (the sedimentation efficiency, $f_{\rm sed}$) controlling the droplet radius ratio for which updrafting, convection-driven wind can no longer keep cloud droplets aloft. This treatment results in cloud droplet abundances that decrease with altitude until subsaturation is reached \citep[see][]{ackemanandmarley2001}.

Thoroughly exploring the moist greenhouse Inner Edge limit(s) of the Habitable Zone requires a series of experiments. The first experiment seeks to constrain the global average temperature threshold for the moist greenhouse climate states from upper atmosphere isotherm temperatures \citep[derived from experiments in][]{kastingetal2015}, and to capture the potentially diminished radiative heating rates above, and convective overshooting from, convective cloud decks \citep{windsoretal2022}. The second experiment tests bulk cloud properties including grids of fractional cloudiness and sedimentation efficiency. The third experiment tests the predictions of the moist greenhouse Inner Edge limits from the climate model in inverse mode (top of atmosphere equilibrium), to that of the forward model (radiative equilibrium) \citep[for detailed descriptions of the inverse and forward models see, e.g.,][]{windsoretal2022}. Finally, planet scale moist greenhouse worlds \citep[or tropical planets,][]{pierrehumbert2011} are unfamiliar to the Solar System and likely have distinctive spectral features. We use a high fidelity line-by-line radiative transfer tool to model characteristic reflectance and transit transmission spectra of moist greenhouse worlds.

\subsection{Determining Moist Greenhouse Temperatures}

The global average surface temperature of the moist greenhouse climate states is determined following methods similar to \citet{kastingetal1993}. We use the climate model from \citet{windsoretal2022} in inverse mode and adopt an Earth-like surface gravity of 9.8\,m\,s$^{-2}$, a non-condensible gas surface pressure of 1\,bar, atmospheric mixing ratios of 0.99 for \ch{N2} and 0.01 for \ch{Ar}, and a condensible\,---\,but well-mixed\,---\,volume mixing ratio of 360\,ppm for \ch{CO2}, but diverge slightly from the albedo-tuned models of \citet{kastingetal1993} by prescribing a gray ocean world planetary surface albedo of 0.06 \citep{windsoretal2022}. To determine the surface temperature where the onset of a moist greenhouse state occurs, we follow \citet{kastingetal1993} and perform calculations at 1\,AU from a Sun-like host, with an incident flux of 1,360\,W\,m$^{-2}$. We initialize two grids of inverse mode climate models with diverging upper atmosphere isothermal temperatures of 150~K \citep{kastingetal2015} and 200~K, while maintaining the same surface temperatures starting at 288.5\,K and incrementally increase to 430\,K in 1\,K steps. We then select where\,---\,in surface temperature space\,---\,the stratospheric \ch{H2O} volume mixing ratio exceeds 1$\times 10^{-3}$ g/g\,---\,signifying the moist greenhouse climate state criterion \citep{kasting1988,kastingetal1993,kopparapu2013}\,---\,and adopt these associated surface temperatures (for each of the two explored stratospheric isotherms) moving forward.

\subsection{Determining Moist Greenhouse Inner Edge Boundaries}

After determining the maximum surface temperatures that result in moist greenhouse climate states for terrestrial planets around a Sun-like star, we adopt this surface temperature(s), and associated upper atmospheric isotherm, to recalculate the moist greenhouse Habitable Zone for the stellar main sequence, following similar methods to earlier works \citep{kastingetal1993,kopparapu2013,kopparapuetal2014}. Our Inner Edge of the Habitable Zone calculations consider main sequence spectral types with stellar effective temperatures in the range of 2,600\,K to 7,200\,K which encompasses the range of F,G,K, and M hosts. We adopt similar stellar models to \citet{kopparapu2013,kopparapuetal2014}, leveraging the ``BT\_Settl'' grid of simulations  \citep{allardetal2003,allardetal2007}. Upon model initialization, the total energy flux of the incident starlight is interpolated over the 38 shortwave spectral bins of the climate model and is normalized to Modern Earth's incident flux of 1,360\,W\,m$^{-2}$. Further, we sample the stellar host grid by selecting stellar effective temperatures in steps of 200~K, and adopt a constant Sun-like metallicity. To capture a wide range of potential cloud properties arising on moist greenhouse worlds, we adopt a logarithmic grid of $f_{\rm sed}$ values ranging from an inefficient value of 0.01, to an efficient value of 10. Finally, to capture the influence of cloud fraction on the moist greenhouse Inner Edge calculations, we explore fractional cloudiness ranging from 0\% (no clouds) to 100\% (full clouds) in steps of 25\%.



\subsection{Forward Model Climates}


 
 Inverse modeling ultimately estimates the amount of incident flux required to maintain specific climate states according to host spectral type as a function of climate-relevant atmospheric and planetary properties. To test the reliability of the inverse climate model and its predictions for the moist greenhouse Habitable Zone, we run a suite of climate models in ``forward'' mode.  Overall, we sample a grid of 6 moist greenhouse forward models from select representative stellar spectral types (M5V, K7V, K4V, G2V, F6V) from the BT\_Settl grid of simulations \citep{allardetal2003,allardetal2007}, with a fixed global cloud coverage of 100\%, and a sedimentation efficiency of 0.64, which yields droplet mean effective radii characteristic of Modern Earth.


The climate model's forward mode fixes the top-of-atmosphere incident stellar flux and iteratively steps towards an equilibrium solution by computing and applying layer-dependant radiative heating rates  \citep[see][]{windsoretal2022}. In this scheme, any model layers that achieve thermal gradients above a prescribed pseudo-adiabat \citep{kasting1988,kastingetal1993} are relaxed back onto this adiabatic structure \citep{kasting1988,kastingetal1993,windsoretal2022}. Each iterative timestep omits energy conservation to speed up model convergence through allowing model layers to convectively adjust instantaneously. Model convergence is achieved when the net longwave flux and the net shortwave flux at the top-of-atmosphere achieve 0.001\% similarity. To ensure maximum model stability in the forward mode and a stable solution, we adopt at least 400 iterative timesteps \citep{windsoretal2022}.


\subsection{Coupled Spectral Models}

Coupled climate-to-spectral models are invaluable tools for exoplanet characterization studies \citep{wordsworthetal2022}. Further, the importance of the moist greenhouse climate state\,---\,in terms of planetary habitability\,---\,advocates for further theoretical and observational study. As in \citet{windsoretal2022}, we pair our climate model to the Spectral Mapping Atmospheric Radiative Transfer (\texttt{SMART}) model \citep{meadowsandcrisp1996,robinson2017} to efficiently generate high fidelity reflectance (0.3--4.5\,{\textmu}m) and transmission (0.1--10\,{\textmu}m) spectral models for each generated climate state. Finally, we explore synthetic observations for a moist greenhouse Earth-sized world based on \textit{JWST} results from \citet{lustig-yaegeretal2023}.


\section{Results}\label{sec:Results}

\begin{figure}
     \centering
         \centering
         \includegraphics{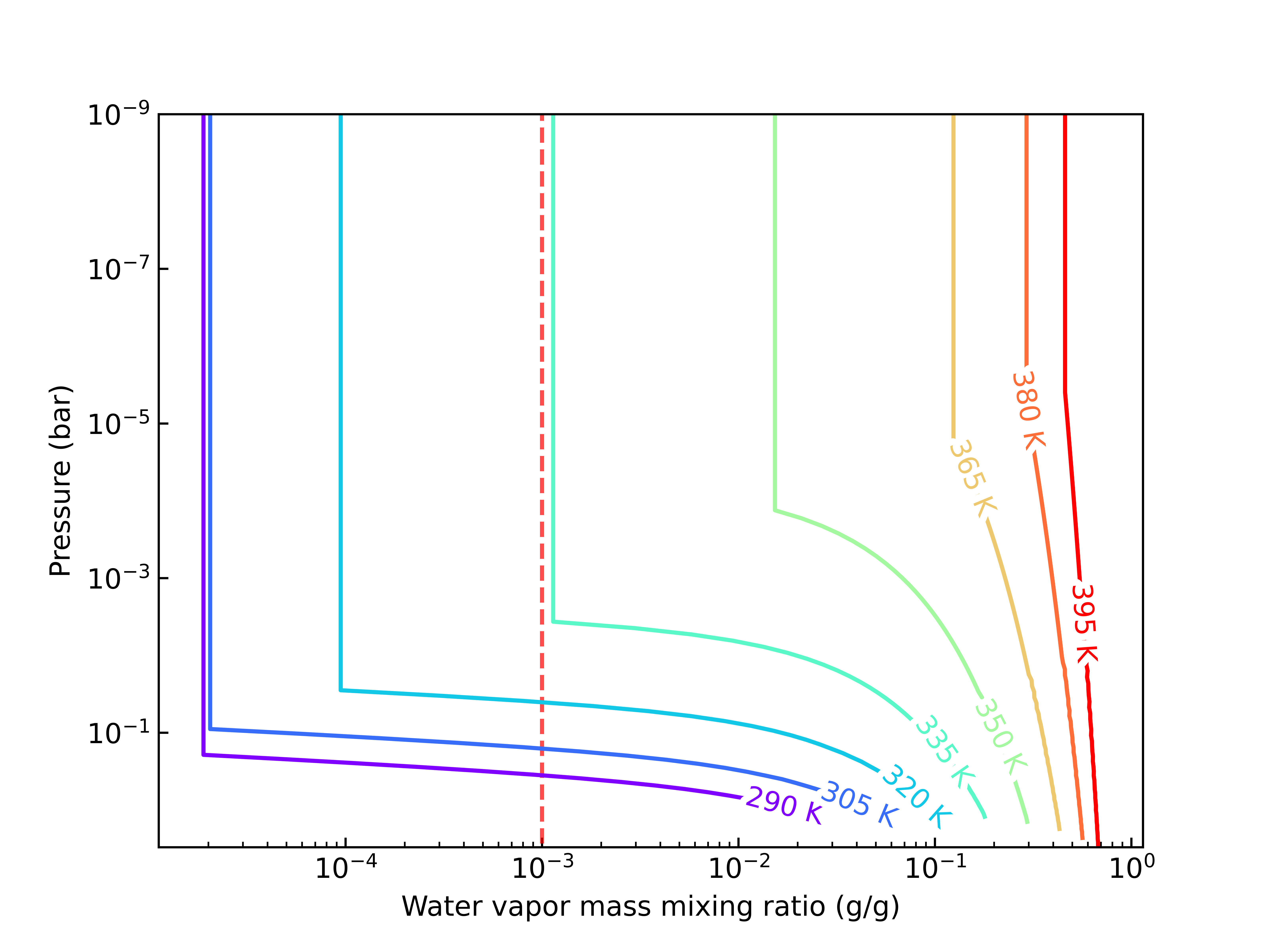}
         \caption{Water vapor volume mixing ratios for climate states in 15\,K increments for the standard 200\,K stratospheric isotherm case. The dashed vertical line in red denotes where the mixing ratios cross the threshold of $10^{-3}$ for the moist greenhouse state. The transition to a moist greenhouse climate state occurs at a surface temperature of 333\,K.}
         \label{fig:water_vapor_mmr_200k}
\end{figure}
\begin{figure}

         \centering
         \includegraphics{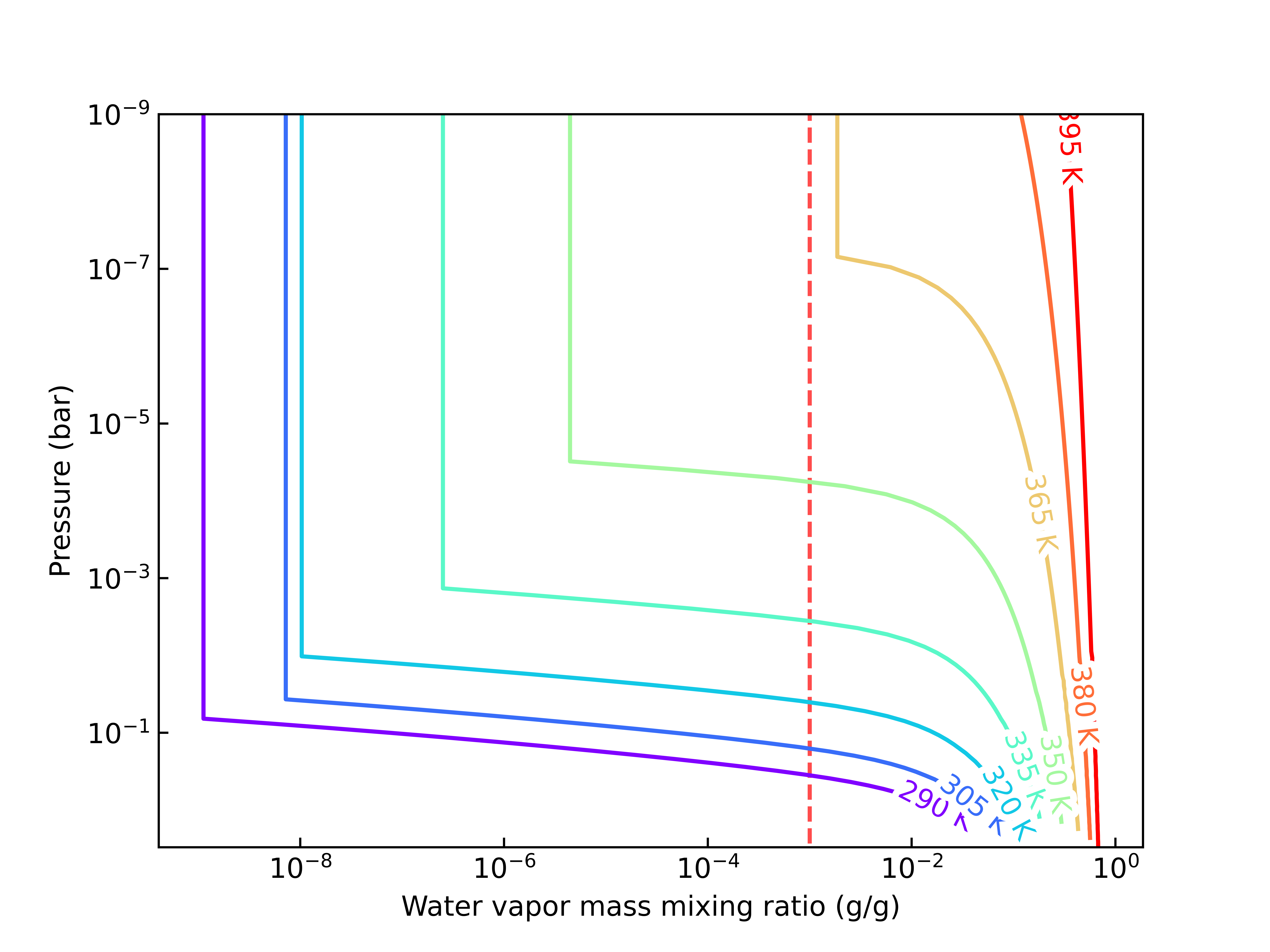}
         \caption{Water vapor volume mixing ratios for climate states in 15\,K increments for the 150\,K stratospheric isotherm case. The dashed vertical line in red denotes where the mixing ratios cross the threshold of $10^{-3}$  for the moist greenhouse state. In these climate models the moist greenhouse climate state is associated with a surface temperature of 363\,K.}
         \label{fig:water_vapor_mmr_150k}

\end{figure}

Figures \ref{fig:water_vapor_mmr_200k} and \ref{fig:water_vapor_mmr_150k} show our modeled sensitivities to the assumed stratospheric isotherm for cloudy moist greenhouse climate states. Some previous studies considered the strong temperature dependence of the atmospheric cold trap by exploring different upper atmosphere isotherm temperatures \citep{kastingetal2015}. \citet{kastingetal2015} suggest adopting an upper atmospheric isotherm of 150\,K due to the lack of ozone absorption in the stratospheric regions of oxygen-free models  \citep[e.g.,][]{kastingetal1993,selsisetal2007,kopparapuetal2013,kopparapuetal2014}. We explore sensitivity to the adopted stratospheric temperature by exploring isotherms at 200\,K \citep{kasting1988,kastingetal1993,selsisetal2007,kopparapu2013,kopparapuetal2014} and 150\,K \citep{kastingetal2015}. 

For the inverse mode climate models where a 200\,K isotherm is adopted in the upper atmosphere, the surface temperature that results in an upper atmosphere water vapor mixing ratio above 10$^{-3}$ is 333\,K. This surface temperature is close to the value of 340\,K reported in previous works for an Earth-sized world \citep{kastingetal1993,kopparapuetal2013}. Conversely, the 150\,K isothermal upper atmosphere climates (Figure \ref{fig:water_vapor_mmr_150k}) result in a transition to a moist greenhouse state at a surface temperature of 363\,K\,---\,30\,K higher than predicted by the warmer stratospheric isotherm models.




The climate model scheme that is responsible for the cold-trapping mechanism is strongly dependent on the stratospheric isotherm temperature. The difference in surface temperatures for a transition to a moist greenhouse state is a direct result of this mechanism, and the more-simplified approach to radiative transfer makes it difficult to determine if a stratosphere as cold as 150\,K would arise using the forward climate model. 

We further assessed the stratospheric thermal structure using the more-sophisticated radiative transfer treatments in \texttt{SMART} \citep{meadowsandcrisp1996}. Inspection of the resulting net radiative heating rates for partially clouded moist greenhouse worlds favor adopting the warmer isotherm of 200\,K. Finally, adopting the warmer isotherm minimizes the orbital distance from the host star for Inner Edge considerations, suggesting that further study is needed to assess a more optimal upper atmosphere isothermal temperature. Nevertheless, we take the moist greenhouse transition surface temperature of 363\,K as an upper limit, and simultaneously compute moist greenhouse Habitable Zone boundaries for both stratospheric thermal profiles to account for uncertainty in the stratospheric temperature.

\subsection{Moist Greenhouse Cloudy Inner Edge Habitable Zone}

\begin{figure}
    \centering
    \includegraphics[width=\textwidth,height=\textheight,keepaspectratio]{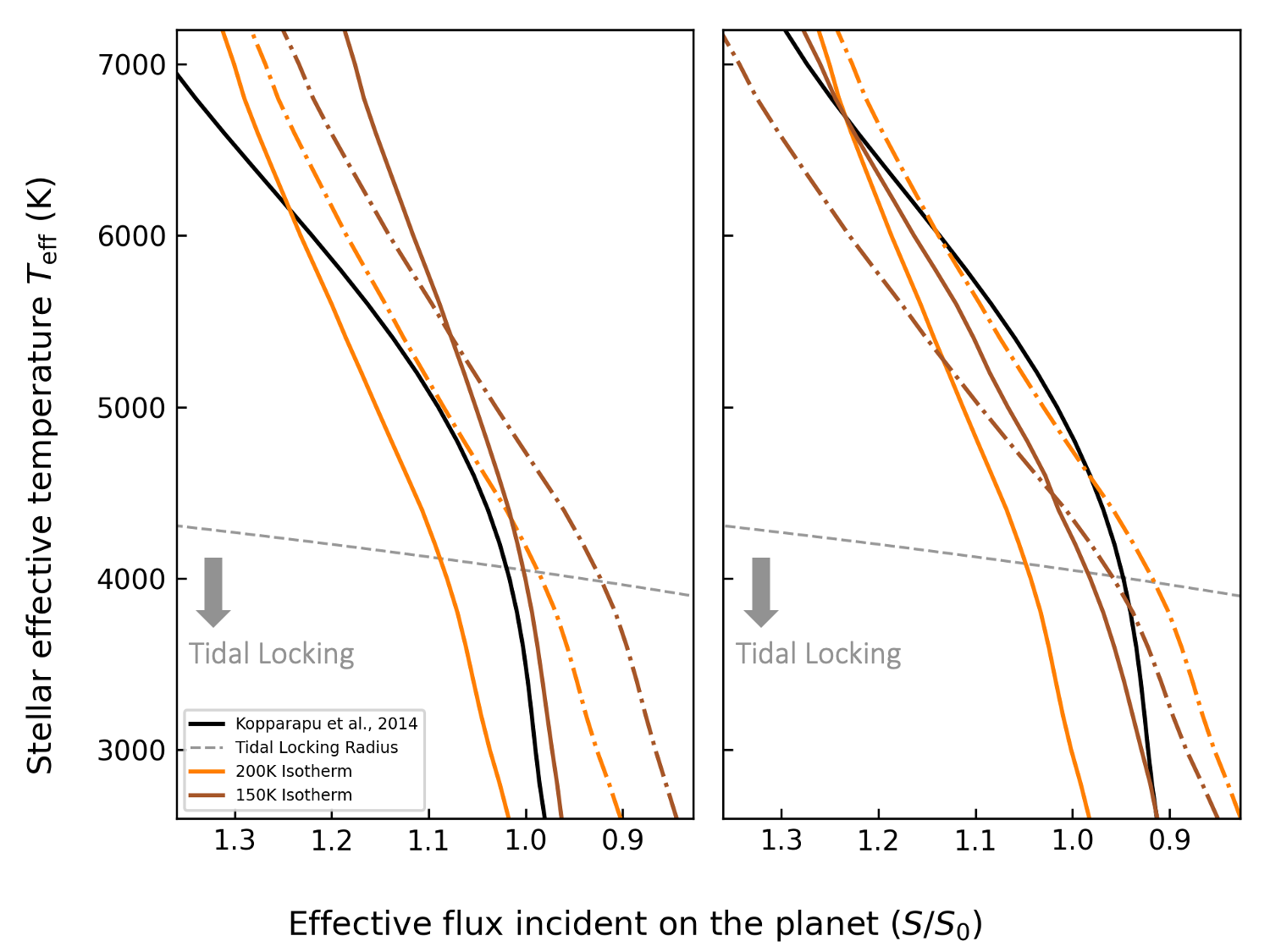}
    \caption{Moist greenhouse Habitable Zone boundaries for super-Earth worlds (left) and Earthsize worlds (right) with 50\% cloud cover and a 200\,K stratosphere (orange) versus a 150\,K stratosphere (brown). For these boundaries, solid lines correspond to Inner Edge climates with a very low cloud sedimentation efficiency (0.01) while the dash-dot lines correspond to Inner Edge climates with an intermediate sedimentation efficiency (1.3). For comparison, cloud-free runaway greenhouse inner edge results from  \citet{kopparapuetal2014} are shown as a black solid line. The tidal locking distance is also shown (dashed).}

    \label{fig:50C_MGIHZ.png}
\end{figure}

\begin{figure}
    \centering
    \includegraphics[width=\textwidth,height=\textheight,keepaspectratio]{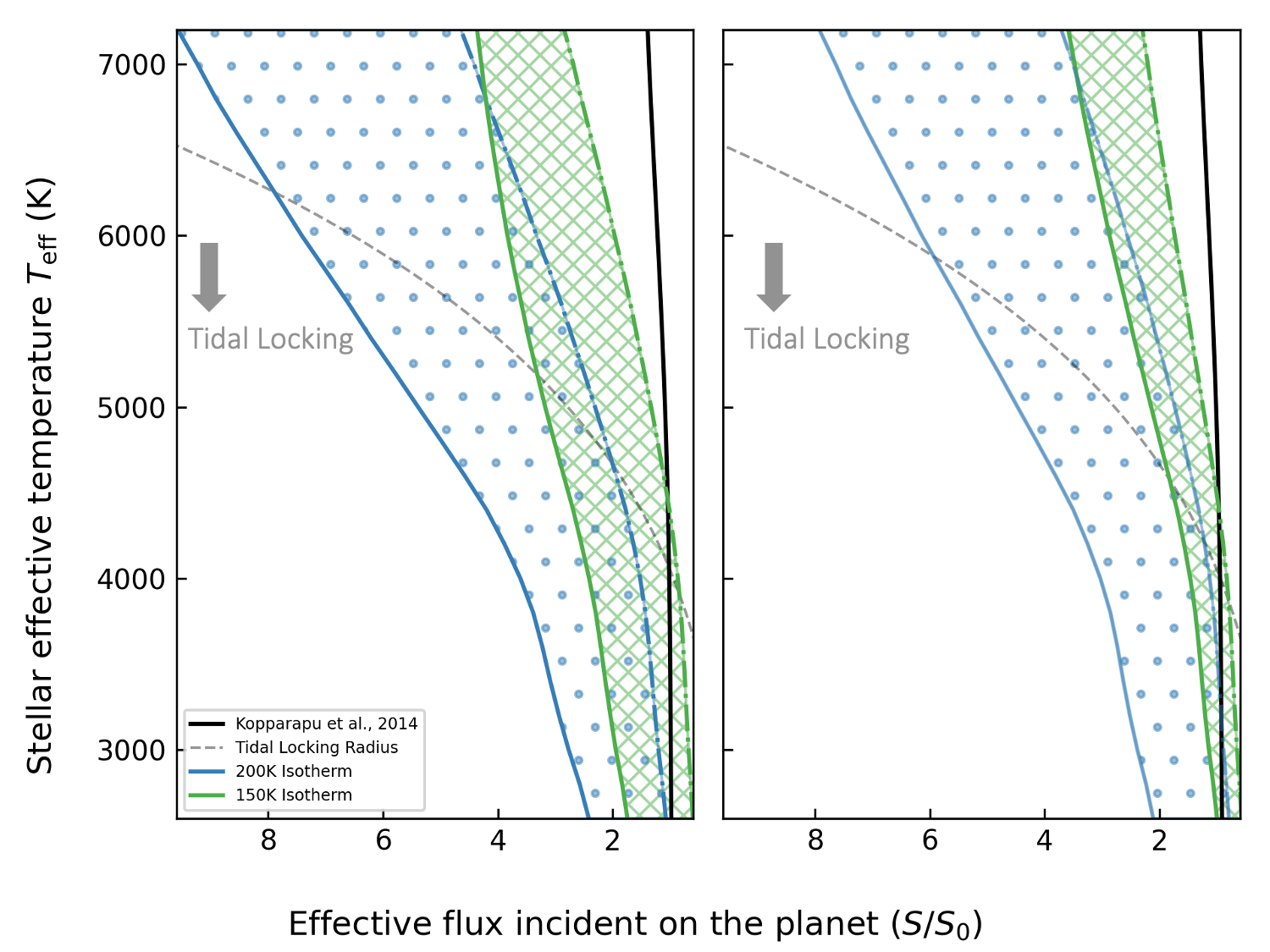}
    \caption{Moist greenhouse Habitable Zone boundaries for super-Earth (left) and Earth-sized worlds (right) with 100\% cloud cover and a 200\,K stratosphere (blue) versus a 150\,K stratosphere (green). Solid lines correspond to Inner Edge climates with a very low cloud sedimentation efficiency (0.01), while dash-dot lines correspond to Inner Edge climates with an intermediate sedimentation efficiency (1.3). For comparison, cloud-free runaway greenhouse inner edge results from \citet{kopparapuetal2014} are shown (black solid line). The tidal locking distance is also shown (dashed line).}

    \label{fig:100C_MGIHZ.png}
\end{figure}

Figures \ref{fig:50C_MGIHZ.png} and \ref{fig:100C_MGIHZ.png} show the location(s) of the cloudy moist greenhouse driven Inner Edge limit(s) of the Habitable Zone, as a function of the stellar effective temperature, stellar insolation(s), fractional cloudiness, and sedimentation efficiency. The resulting cloudy moist greenhouse boundaries are compared directly to key earlier results \citep{kopparapu2013,kopparapuetal2014}. Note the strong dependence on isothermal upper atmosphere temperature (solid colored lines) that arises due to its influence on cloud structure, and, thus, cloud greenhouse effect. For all tested spectral types, cooler upper atmosphere isotherms result in \textit{cloud-free} moist greenhouse limits located at as much as 200\% higher instellations than in earlier results \citep{kastingetal1993,kopparapuetal2013,kopparapuetal2014}; however, the clouded Inner Edge limits show the opposite effect due to cloud top temperature-driven greenhouse effects. 

Across all tested thermal profiles, sedimentation efficiencies, and stellar hosts, the climates that result in the highest planetary Bond albedo are the climates in which the planet is completely cloud covered. For instance, around Sun-like stars ($T_{\rm eff}$ = 5800\,K), 100\% clouded super-Earth climates achieve a moist greenhouse boundary at roughly 7 times the current instellation of Earth in the most inefficient sedimentation case, and 1.04 times the current instellation of Earth for 25\% clouded and high sedimentation efficiency (see Table~\ref{tab:sunlike_MGIHZ}). For comparison, \citet{kopparapuetal2014} reported a runaway greenhouse limit at an instellation of 1.19\,$S_{\rm{eff}}$ for a super-Earth sized planet around the Sun. 

The moist greenhouse limit for super-Earths in comparison to Earth-sized planets indicates a strong dependence on planetary mass and background \ch{N2} abundance. For Sun-like stars, the 100\% clouded Earth-sized climate state with inefficient sedimentation ($f_{\rm sed}$ = 0.01) results in an instellation limit of 5.7 $S_{\rm{eff}}$ versus 6.9\,$S_{\rm{eff}}$ for the super-Earth (R$_{\rm p}$ = 1.5 Earth radius) with similar cloud parameters, which indicates moist greenhouse climate states at instellations 1.2 times higher. Similarly, the instellation limits for the 50\% clouded climates with a sedimentation efficiency ($f_{\rm sed}$ = 1.3) result in an instellation limit of 1.11\,$S_{\rm{eff}}$ for an Earth-sized planet, versus 1.16\,$S_{\rm{eff}}$ for a super-Earth with a Sun-like host. The 50\% clouded cases in particular are comparable to the instellation limits of the runaway greenhouse climate states reported in \citet{kopparapuetal2014}, with a difference in super-Earth to Earth-sized planets' instellation range of 0.08\,$S_{\rm{eff}}$. Earlier models \citep{kastingetal1993,selsisetal2007,kopparapuetal2013,kopparapuetal2014} adopted a tuned planetary surface albedo of roughly 0.3 to account for the cloud albedo effect while our models adopt an ocean world planetary surface albedo of 0.06. Adopting Earth's true planetary surface albedo of 0.13 likely results in the 50\% clouded moist greenhouse limits surpassing those of \citet{kopparapuetal2013,kopparapuetal2014} in all cases.

\begin{figure}
    \centering
    \includegraphics{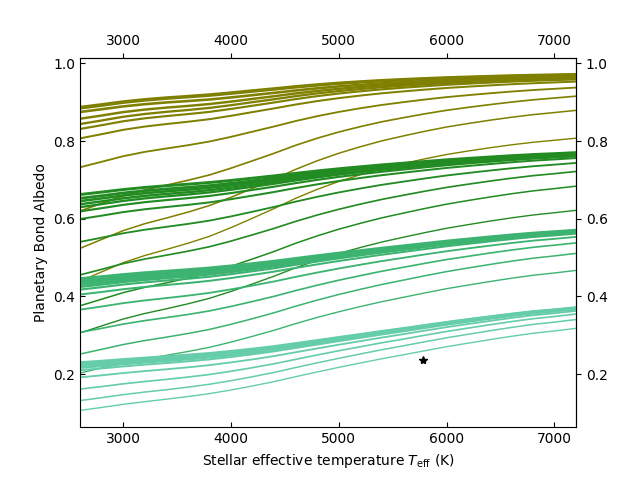}
    \caption{Planetary Bond albedo as a function of stellar effective temperature for super-Earth climates with an upper atmosphere isotherm of 200\,K. The black star marker corresponds to the Bond albedo from \citet{kopparapuetal2014} (their Figure 2) for a planet with a surface temperature of 333\,K. Line thickness corresponds to cloud sedimentation efficiency, where the thickest lines are representative of sedimentation efficiencies of 0.01, and the thinnest lines represent sedimentation efficiencies of 10. Each shade of green represents different cloud percentages, with olive green (uppermost) corresponding to 100\% cloudy, forest green representing 75\% cloudy, medium sea green representing 50\% cloudy, and medium aqua marine (lowermost) representing 25\% cloudy. Most notable is the divergence of the clouded planetary Bond albedo with the runaway greenhouse Bond albedo from \citet{kopparapuetal2014}, where the clouded climate models here account for the cloud greenhouse effect instead of just the cloud albedo effect, and can present larger planetary Bond albedo values comparatively.}
    \label{fig:SE_200K_Bond_Albedo}
\end{figure}

Figure ~\ref{fig:SE_200K_Bond_Albedo} shows planetary Bond albedo as a function of stellar effective temperature ($T_{\rm eff}$) for super-Earth sized planets with an upper atmosphere isotherm of 200\,K. Climates with 100\% cloud cover result in the highest planetary Bond albedo values, achieving 0.75--0.91 for super-Earths orbiting the Sun, for the full range of sedimentation efficiencies tested. Climates with 50\% clouds result in a planetary Bond albedo range of 0.40--0.53. Finally, climates with 25\% clouds result in a planetary Bond albedo range of 0.26--0.32. We compare these values directly to \citet{kopparapuetal2014}, where the reported Bond albedo for a saturated troposphere super-Earth climate with a 2.5\,bar \ch{N2} background atmosphere and surface temperature of 333\,K is 0.23, with an effective incident flux of $\approx$1.2. The differences in planetary Bond albedo from 50\% clouded cases compared with the results from \citet{kopparapuetal2014} arise due to the different radiative environments within the climate model. Namely, the cloud greenhouse effect and a slight difference in moist greenhouse onset surface temperature.



\begin{table}
\resizebox{\textwidth}{!}{\begin{tabular}{ccccc}
\hline

$f_{\rm sed}$              & 100\% Cloudy & 75\% Cloudy & 50\% Cloudy & 25\% Cloudy \\ \hline
0.01 (1 $ M_{\rm \oplus}$) & 5.80       & 1.57      & 1.17             & 1.02   \\
0.01 (3.5 $ M_{\rm \oplus}$) & 7.01       & 1.65      & 1.22             & 1.05    \\ \hline
0.02 (1 $ M_{\rm \oplus}$) & 5.40       & 1.56      & 1.17             & 1.02   \\
0.02 (3.5 $ M_{\rm \oplus}$) & 6.72       & 1.63      & 1.21             & 1.05   \\ \hline
0.04 (1 $ M_{\rm \oplus}$) & 5.02       & 1.54      & 1.16             & 1.02   \\
0.04 (3.5 $ M_{\rm \oplus}$) & 5.91       & 1.62      & 1.21             & 1.05   \\ \hline
0.08 (1 $ M_{\rm \oplus}$) & 4.55       & 1.52      & 1.16             & 1.02   \\
0.08 (3.5 $ M_{\rm \oplus}$) & 5.42       & 1.61      & 1.20             & 1.05   \\ \hline
0.16 (1 $ M_{\rm \oplus}$) & 4.00       & 1.49      & 1.15             & 1.01   \\
0.16 (3.5 $ M_{\rm \oplus}$) & 5.06       & 1.60      & 1.20             & 1.05   \\ \hline
0.32 (1 $ M_{\rm \oplus}$) & 3.35       & 1.44      & 1.14             & 1.01   \\
0.32 (3.5 $ M_{\rm \oplus}$) & 4.70       & 1.58      & 1.20             & 1.05   \\ \hline
0.64 (1 $ M_{\rm \oplus}$) & 2.76       & 1.39      & 1.12             & 1.01   \\
0.64 (3.5 $ M_{\rm \oplus}$) & 4.13       & 1.54      & 1.19             & 1.05   \\ \hline
1.2  (1 $ M_{\rm \oplus}$) & 2.38       & 1.36      & 1.11             & 1.00   \\
1.2  (3.5 $ M_{\rm \oplus}$) & 3.11       & 1.46      & 1.17             & 1.04   \\ \hline
2.5  (1 $ M_{\rm \oplus}$) & 2.30       & 1.37      & 1.12             & 1.00     \\
2.5  (3.5 $ M_{\rm \oplus}$) & 2.57       & 1.41      & 1.15             & 1.03   \\ \hline
5.0  (1 $ M_{\rm \oplus}$) & 2.11       & 1.35      & 1.12             & 1.00  \\
5.0  (3.5 $ M_{\rm \oplus}$) & 2.40       & 1.42      & 1.16             & 1.04   \\ \hline
10   (1 $ M_{\rm \oplus}$) & 1.78       & 1.28      & 1.10             & 1.00   \\
10   (3.5 $ M_{\rm \oplus}$) & 2.06       & 1.37      & 1.15             & 1.04   \\
\end{tabular}}
\caption{Instellation values for clouded Earth \& super-Earth sized planets around a Sun-like host. Here, this table explores the moist greenhouse Habitable Zone limits as a function of fractional cloudiness ($f_{\rm cld}$) and sedimentation efficiency ($f_{\rm sed}$). }
\label{tab:sunlike_MGIHZ}
\end{table}




Following \citet{kopparapuetal2013,kopparapuetal2014}, we provide parametric equations for fitting the moist greenhouse Inner Edges:

\begin{equation*}
    S_{\rm eff} = S_{\rm eff \odot} + aT_{\rm \star} + bT_{\rm \star}^{\rm 2} + cT_{\rm \star}^{\rm 3} + dT_{\rm \star}^{\rm 4} \ ,
\end{equation*}
where $T_{\rm \star} = T_{\rm eff} - 5780 \rm{K}$ and the coefficients are given in Table~\ref{tab:Fit_coefficients}. The corresponding Inner Edge distances, $d$, can be calculated using \citep{kopparapuetal2013,kopparapuetal2014}:

\begin{equation*}
    d = \left(  \frac{L/L_{\rm \odot}}{S_{\rm eff}}   \right)^{\rm 0.5} \rm{AU} \ ,
\end{equation*}
where $L/L_{\rm \odot}$ is the star luminosity, $L$, relative to that of the Sun, $L_{\rm \odot}$. 

\begin{table}
\resizebox{\textwidth}{!}{\begin{tabular}{ccccccc}
\hline
Constant                                     & Recent Venus                 & Runaway Greenhouse            & 100\% Cloudy, $f_{\rm sed}$ = 0.01 & 100\% Cloudy, $f_{\rm sed}$ = 1.25 & 50\% Cloudy, $f_{\rm sed}$ = 0.01 & 50\% Cloudy, $f_{\rm sed}$ = 1.25 \\ \hline
$S_{\rm eff \odot}$ (1 $M_{\rm \oplus}$) & 1.776                        & 1.107                         & 5.746                              & 2.360                                 & 1.170                             & 1.111                                \\
$S_{\rm eff \odot}$ (3.5 $M_{\rm \oplus}$) & ...                          & 1.188                         & 6.942                              & 3.078                                 & 1.214                             & 1.162                                \\ \hline
a (1 $ M_{\rm \oplus}$)                   & 2.136 $\times 10^{\rm -4}$   & 1.332 $\times 10^{\rm -4}$    & 1.742 $\times 10^{\rm -3}$         & 9.037  $\times 10^{\rm -4}$           & 7.622 $\times 10^{\rm -5}$        & 1.102 $\times 10^{\rm -4}$           \\
a (3.5 $ M_{\rm \oplus}$)                   & ...                          & 1.433 $\times 10^{\rm -4}$    & 2.114 $\times 10^{\rm -3}$         & 1.096  $\times 10^{\rm -3}$           & 8.092 $\times 10^{\rm -5}$        & 1.024 $\times 10^{\rm -4}$           \\ \hline
b (1 $ M_{\rm \oplus}$)                   & 2.533 $\times 10^{\rm -8}$   & 1.158 $\times 10^{\rm -8}$    & 1.752 $\times 10^{\rm -9}$         & 8.464 $\times 10^{\rm -8}$            & -2.063 $\times 10^{\rm -9}$       & -7.222 $\times 10^{\rm -9}$          \\
b (3.5 $ M_{\rm \oplus}$)                   & ...                          & 1.707 $\times 10^{\rm -8}$    & -8.619 $\times 10^{\rm -9}$        & 7.467 $\times 10^{\rm -8}$            & -1.650 $\times 10^{\rm -9}$       & -5.388 $\times 10^{\rm -9}$          \\ \hline
c (1 $ M_{\rm \oplus}$)                   & -1.332 $\times 10^{\rm -11}$ & -8.308  $\times 10^{\rm -12}$ & -1.033 $\times 10^{\rm -10}$       & -3.309 $\times 10^{\rm -11}$          & -4.084 $\times 10^{\rm -12}$      & -4.293 $\times 10^{\rm -12}$         \\
c (3.5 $ M_{\rm \oplus}$)                   & ...                          & -8.968 $\times 10^{\rm -12}$  & -1.224 $\times 10^{\rm -10}$       & -4.696 $\times 10^{\rm -12}$          & -4.316 $\times 10^{\rm -12}$      & -5.127 $\times 10^{\rm -12}$         \\ \hline
d (1 $ M_{\rm \oplus}$)                   & -3.097 $\times 10^{\rm -15}$ & -1.931 $\times 10^{\rm -15}$  & -1.344 $\times 10^{\rm -14}$       & -5.796 $\times 10^{\rm -15}$          & -5.206 $\times 10^{\rm -16}$      & -2.901 $\times 10^{\rm -16}$         \\
d (3.5 $ M_{\rm \oplus}$)                   & ...                          & -2.804 $\times 10^{\rm -15}$  & -1.547 $\times 10^{\rm -14}$       & -7.353 $\times 10^{\rm -15}$          & -5.820 $\times 10^{\rm -16}$      & -3.983 $\times 10^{\rm -16}$         \\ \hline
\end{tabular}}
\caption{Constants for the parametric fits to the partially clouded moist greenhouse Habitable Zone limits. The values for recent Venus and runaway greenhouse are from \citet{kopparapuetal2014} and are included for direct comparison. Only 100\% and 50\% clouded climates are fitted to provide a manageable sample of cloud cover percentages that may arise on rocky worlds. Further, we adopt only the inefficient sedimentation efficiency ($f_{\rm sed}$ = 0.01) and Earth-like sedimentation efficiency \citep[$f_{\rm sed}$ = 1.2; see][]{windsoretal2022}. }
\label{tab:Fit_coefficients}
\end{table}

\subsection{Moist Greenhouse Forward Models}

\begin{figure}
    \centering
    \includegraphics[width=\textwidth,height=\textheight,keepaspectratio]{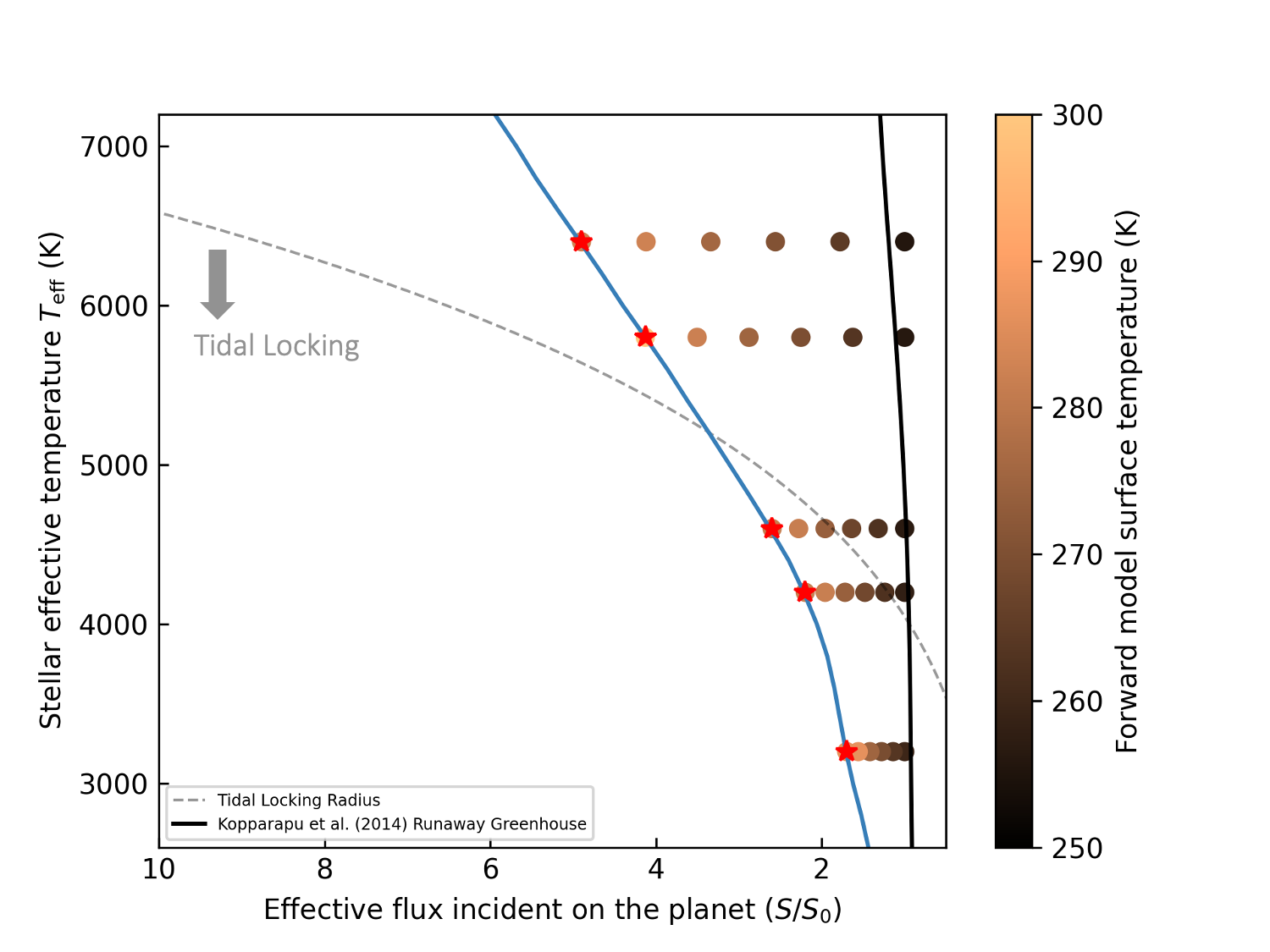}
    \caption{Moist greenhouse Inner Edge forward models. Forward climate models are represented by solid circle markers with shades of copper correlating to the planetary surface temperature. The blue solid line corresponds to a sedimentation efficiency of 0.64 and is analogous to Earth clouds. Red stars represent forward model climates in which the model predicts a runaway climate state. In this case the runaway climate states agree well with the moist greenhouse Inner Edge predictions by the Inverse Models.}
    \label{fig:MG_forward}
\end{figure}

We verify the Inner Edge Limit results of the inverse model through direct comparison to super-Earth forward models. Instead of assuming an atmospheric thermal structure like the inverse model, the forward model adopts a fixed top-of-atmosphere incident flux and iterates towards an equilibrium solution following convective adjustment methods \citep{kasting&ackerman1986,windsoretal2022}. However, instead of modeling the entire inner edge of the Habitable Zone we select a sub-grid of characteristic clouded climates, as this climate regime is close to an unstable runaway greenhouse state in which the forward model encounters computational instability. Each forward model is initialized with a surface temperature of 288\,K, adopts a fractional cloudiness of 100\% and sedimentation efficiency of 0.64, and explores 6 linearly spaced grid points in incident flux space. Here, the minimum incident flux is 1.0\,$S_{\rm{eff}}$ and the maximum incident flux is extracted from Inner Edge calculations of the previous sections. Figure \ref{fig:MG_forward} showcases the resulting subset of forward models in direct comparison to inverse moist greenhouse calculations. Forward models receiving Modern Earth  levels of incident flux achieve surface temperatures close to 260\,K (which is colder than true Modern Earth due to being fully cloud covered). Forward model climate states remain stable and converged over a range of increased incident flux values. However, the forward model cases that approach the moist greenhouse limit determined by the inverse model become unstable and eventually tip into a ``runaway climate'' state before converging. 

In each unstable forward model case, the surface temperature increases until it reaches temperatures above 400\,K. While this is occurring, cloud feedbacks lead to thicker and thicker cloud decks until enough tropospheric water vapor eventually drives substantial thermal radiative heating  while simultaneously increasing shortwave absorption -- at this point the convective-adjustment scheme in the climate model shuts down completely. These unstable states driven simultaneously by cloud and liquid water vapor feedbacks likely approach regimes where the governing equations of the climate model are no longer physical. In this regime, numerical instabilities in the climate model lead to unphysical runaway model states. This doesn't negate the predictions made by the inverse climate models, however, as climates within these limits tend to behave well in forward mode.

\subsection{Moist Greenhouse Spectral Models}

In this study, we generate high-fidelity, noise-free transit transmission and reflection spectra using \texttt{SMART} \citep{meadowsandcrisp1996,robinson2017} for adaptability to a wide range of observational scenarios. We highlight inverse mode moist greenhouse super-Earth climate states, as these worlds are fairly characteristic of cases across the full range of host star types. Each presented inverse climate model adopts a characteristic thermal profile, atmospheric volume mixing ratios, and global cloud fraction. While stellar effective temperature, stellar radius, orbital distance, and sedimentation efficiency are adopted as free parameters. In order to probe spectral features arising from different cloud sedimentation efficiencies (secondary to global cloud fraction), we select cloud sedimentation efficiency end cases of $f_{\rm{sed}}$ = 0.01, 0.64, and 10.0, which signify sedimentation efficiencies that are inefficient, tuned-to-Earth, and extremely efficient, respectively. Finally, we showcase spectral models in terms of effective transit altitude. With this scheme, most spectral differences arise due to changes in cloud sedimentation efficiency. 
 
\begin{figure}
    \centering
    \includegraphics{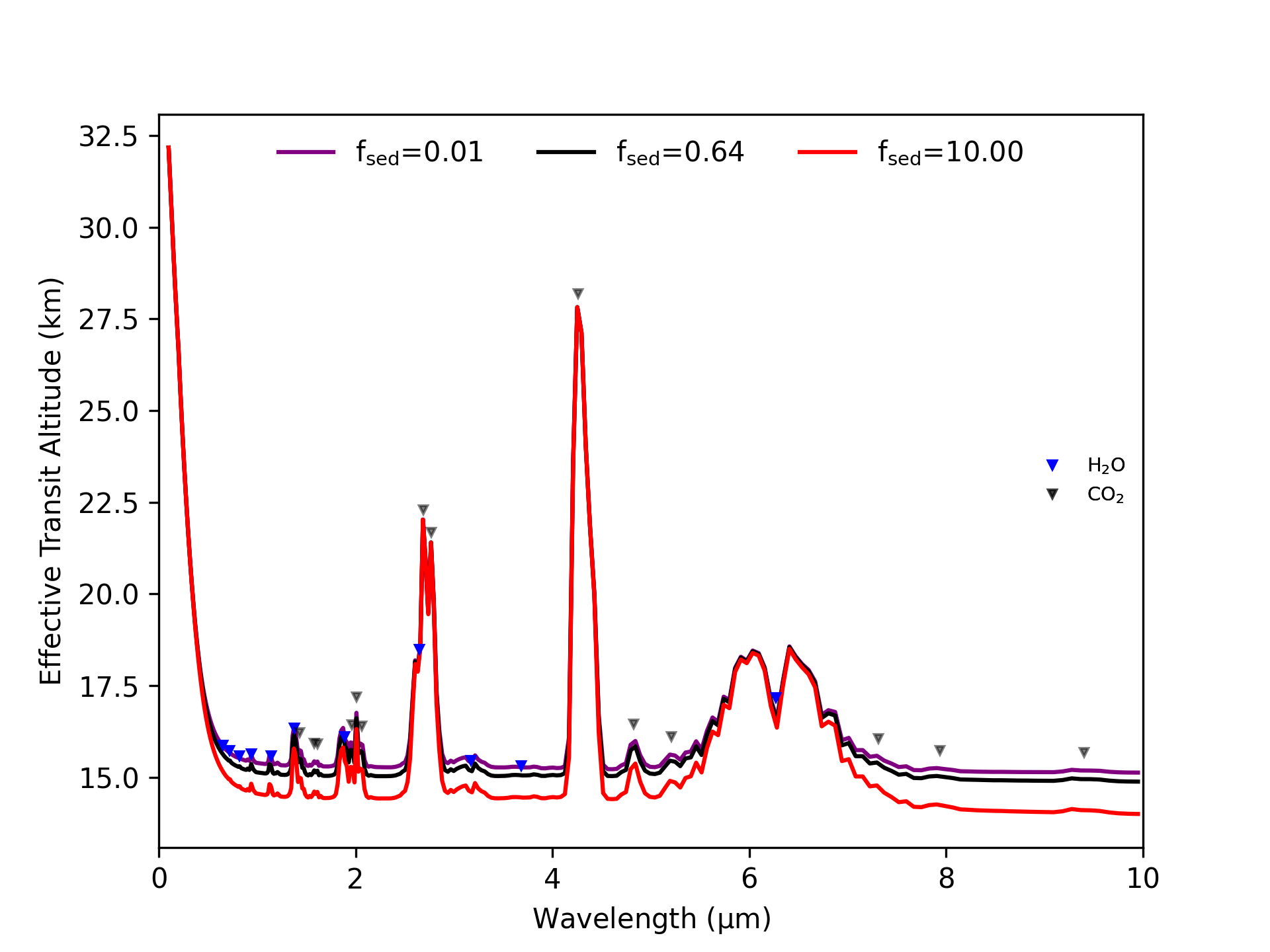}
    \caption{Transit transmission spectra for 100\% clouded moist greenhouse Inner Edge super-Earth worlds around a main sequence K4V with sedimentation efficiency of 0.01 (purple), 0.64 (black), and 10.0 (red). In all cases \ch{H2O} and \ch{CO2} are prominent features above the opaque cloud decks. Secondary characteristics arise where continuum altitude is related directly to cloud deck altitude in the climate models. The trace amount of \ch{CO2} (360\,ppm) yields large transmission features. The high effective transit altitude of the large 6.3\,\textmu m \ch{H2O} feature is indicative of a moist greenhouse state in which the atmosphere above the cloud deck is saturated \citep{kopparapuetal2017}.}
    \label{fig:transit_inverse}
\end{figure}

For the 100\% cloudy model cases explored here, the most apparent features in the transit transmission spectrum models are the \ch{CO2} absorption bands at 2.72 and 4.3\,\textmu m. The \ch{CO2} absorption peak reaches an effective transit altitude greater than 25 kilometers, while the continuum from the intermediate effective transit altitude cloud deck is located around 14 kilometers. The prominence of these \ch{CO2} absorption features\,---\,though it is a trace species in the atmosphere\,---\,emphasizes their importance in characterizing secondary atmospheres of terrestrial planets. Apparent from Figure~\ref{fig:transit_inverse}, changing the height of the cloud deck has little effect on the effective transit altitude of the \ch{CO2} absorption band at 4.3 \textmu m, so that the impact of sedimentation efficiency on cloud deck height may result in a relationship between 4.3 \textmu m \ch{CO2} absorption prominence and sedimentation efficiency.

\begin{figure}
    \centering
    \includegraphics{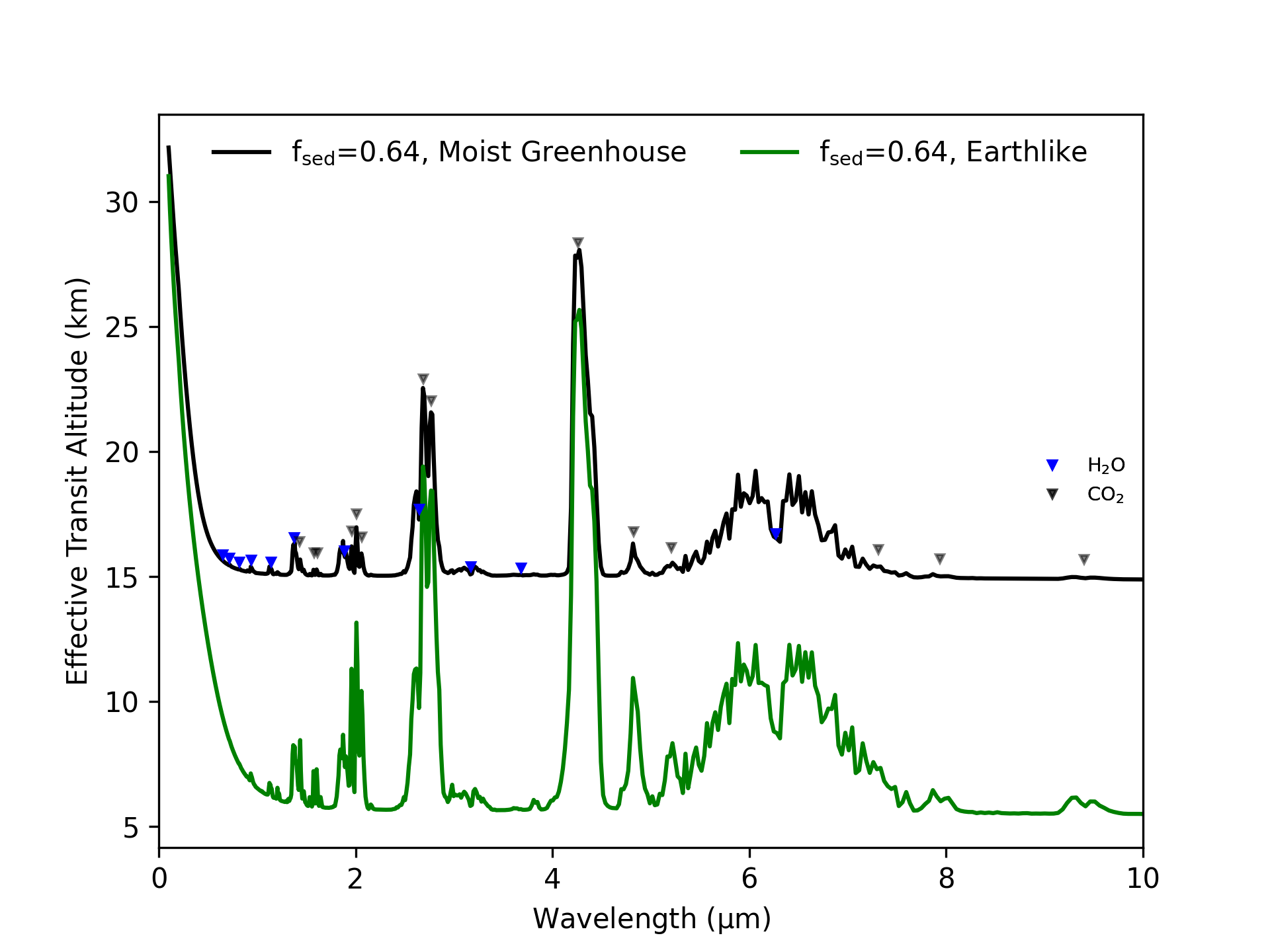}
    \caption{Transit transmission spectra for 100\% clouded moist greenhouse Inner Edge Worlds with a fixed sedimentation efficiency of 0.64 (black line). The green solid line represents a clement climate case, in which we adopt an Earth-like surface temperature of 288\,K and a sedimentation efficiency of 0.64. Despite a much lower volume mixing ratio of water vapor in the upper atmosphere, the amplitude of the 6.3\,\textmu m \ch{H2O} absorption is slightly increased relative to the continuum for the more Earth-like model. }
    \label{fig:clement_inverse}
\end{figure}

The second most prominent spectral feature is the 6.3\,\textmu m \ch{H2O} absorption band, with a peak effective transit altitude of 19\,km. The GCM-derived spectra in \citet{kopparapuetal2017} showcase variability in the 6.3\,\textmu m \ch{H2O} absorption band as a function of planetary rotation rate. Slowly rotating worlds tend to develop strong moist convection in the substellar point regions, whereas rapidly rotating planets tend to not convectively mix water vapor as readily into the upper atmosphere. Our one-dimensional radiative-convective climate model does not resolve processes driven by planetary rotation rate, but we can approximate differences in the spectral features from this process by comparing the spectral features of planets that do not have \ch{H2O} saturated stratospheres with planets in a moist greenhouse state. In the moist greenhouse models, the \ch{H2O} volume mixing ratio is 10$^{\rm{-3}}$, whereas the more clement climate states\,---\,representative of a \textit{fast-rotator}\,---\,achieves \ch{H2O} cold-trap mixing ratios on the order of 10$^{\rm{-6}}$. \citet{kopparapuetal2017} assert that the large difference in the water vapor volume mixing ratio is due to strong convection originating from the substellar point in the permanent dayside in the tidally locked regime. However, our clement climate spectral models in figure \ref{fig:clement_inverse} showcase an increased amplitude of the broad \ch{H2O} absorption feature relative to the cloud-deck continuum when compared to the moist greenhouse climate state. 

\begin{figure}
    \centering
    \includegraphics{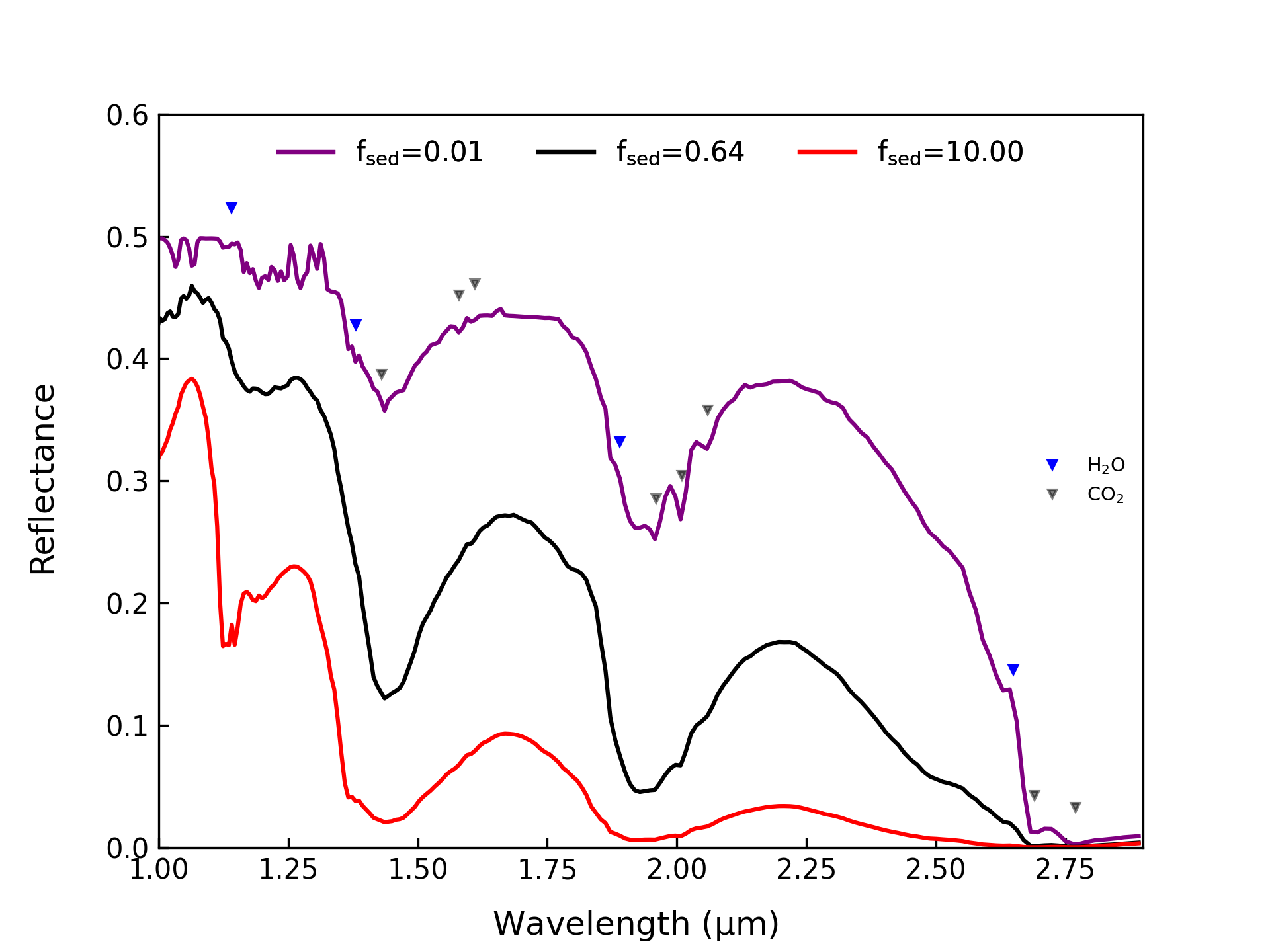}
    \caption{Near-infrared reflection spectra of  100\% clouded moist greenhouse Inner Edge worlds with sedimentation efficiency of 0.01 (purple), 0.64 (black), and 10.0 (red). In call cases, absorption is driven by \ch{H2O} and \ch{CO2}, with \ch{H2O} absorption bands labeled with blue markers, and \ch{CO2} absorption features labeled with grey markers. Sedimentation efficiency is strongly related to spectral reflectance in the 1.0--3.0\,\textmu m region. Here, the climate model with the lowest sedimentation efficiency ($f_{\rm sed}$ = 0.01) results in very vertically-extended cloud decks with small mean effective cloud droplet radii and high albedo.  }
    \label{fig:reflection}
\end{figure}

Figure \ref{fig:reflection} shows reflectance spectra of three inverse mode moist greenhouse climate states. Here, like in Figure~\ref{fig:transit_inverse}, we explore a broad range of sedimentation efficiencies, from an inefficient value of $f_{\rm sed}$ = 0.01, to an efficient value of $f_{\rm sed}$ = 10.0. The highly reflective clouds arising in the $f_{\rm sed}$ = 0.01 case result in reflectance spectrum near infrared continuum values on the order of 0.4 for wavelength ranges bounded by 1.5 and 1.7\,\textmu m, whereas the $f_{\rm sed}$ = 10.0 case results in values of 0.07 in a similar wavelength range. This significant difference in reflectivity is striking and can indicate key wavelength ranges for remote sensing of cloud properties.


\section{Discussion}\label{sec:Discussion}

Convective clouds significantly change the limits of the Inner Edge of the Habitable Zone. In the most extreme case, completely covering the planet with clouds results in a predicted Inner Edge limit of 700 -- 200\% the effective flux Earth receives ($S_{\rm eff}$ = 1.0) and significantly pushes the Inner Edge limit for the Habitable Zone closer to the host star (e.g., 0.38--0.71 AU for a Sun-like star, thus encompassing Venus' orbit at 0.72 AU). Some previous Inner Edge calculations place Earth precariously close to the edge of habitability. For instance, cloud-free models, such as \citet{kastingetal1993,kopparapuetal2013,kopparapuetal2014}, compute the Inner Edge at 0.95, and 0.97 AU, respectively, from the Sun ($S_{\rm eff}$ = 1.11, 1.06), an equivalent increase of just 10-- 6\% of Earth's insolation. These previous studies parameterize the albedo effect of Earth clouds only through tuning the surface albedo of the climate model to mimic that of the Modern Earth. Water clouds arising on terrestrial planets are subject to net warming and net cooling cloud radiative feedbacks \citep{windsoretal2022}, while moist greenhouse climates tend to generate clouds that have a substantial net cooling effect. For instance, our climate models adopt a mean surface albedo of 0.06 and even small cloud fractions (50-25\%) result in a Inner Edge limit that is closer to the host star ($S_{\rm eff}$ = 1.17, 1.02), than for a cloud free ocean world ($S_{\rm eff}$ = 0.94) for Sun-like stars.  

In general, other previous Inner Edge calculations showcase clouds as insulators for planetary habitability from increases in insolation. However, not all cloud parameterizations are equivalent. For instance, \citet{kitzmannetal2010} adopt mean Earth cloud bulk microphysical properties and compute the clouded Inner Edge of the Habitable Zone at a distance of 0.97 AU for a Sun-like star. Other models, such as \citet{selsisetal2007}, highlight the hot, high water content atmospheric states of Inner Edge worlds and prescribe a cloud layer between 0.1 and 1.0 bars in a steam atmosphere with 100 bars of \ch{H2O} and a global average surface temperature of 373~K and compute an Inner Edge of the Habitable Zone limit at 0.46 AU ($S_{\rm eff}$ = 4.73) from the Sun. Here, both studies prescribe a fractional cloudiness of 100\% in order to maximize negative cloud radiative forcing. Equivalent cloud fractions from our study yield Inner Edge limits ranging from $S_{\rm eff}$ = 5.8 (0.42 AU) to $S_{\rm eff}$ = 1.78 (0.75 AU). 

Most information about the role of cloud spatial distribution on the edges of the Habitable Zone has arisen from leveraging three-dimensional global circulation models \citep{yangetal2014,kopparapuetal2016,kopparapuetal2017}, in which bulk cloud properties (e.g., column fractional cloudiness) are more self-consistently computed. \citet{kopparapuetal2017} showcases that high altitude, optically thick clouds forming on the substellar point of rotationally locked worlds help to stabilize climate-cloud radiative feedbacks and can push the Inner Edge of the Habitable Zone $\sim$ 50\% closer from the star for late K dwarf spectral types as compared to the results of \citep{kastingetal1993,kopparapu2013,kopparapuetal2014}. These limits are 80 -- 60\% further from the stellar host than reported from \citet{yangetal2014,kopparapuetal2016}, likely due to the updated water vapor absorption data from \citep{HITRAN2012}. Further, these studies show that there is abundant cloud formation on the permanently irradiated sides of close-in worlds \citep{yangetal2014,wolfeandtoon2014,kopparapuetal2017}. However, some studies also show abundant cloud cover on a permanent night side \citep{kopparapuetal2017} -- demonstrating, in all cases, warm worlds close to the Inner Edge limit of the Habitable Zone present with high fractional cloudiness. As our one-dimensional model is incapable of modeling inherently three-dimensional circulation patterns, we approximate the cloud radiative effects of moist Inner Edge worlds via adopting 100\% cloud cover in our one-dimensional radiative-convective climate model.

Different cloud parameterizations (e.g., extremely efficient sedimentation; $f_{\rm sed}$ = \,10) or changes in fractional cloudiness result in vastly different estimates for the Inner Edge limit orbital distances (see Table~\ref{tab:sunlike_MGIHZ}). Many of the bulk cloud microphysical properties in previous cloudy Inner Edge studies are either unreported \cite{selsisetal2007,yangetal2014} or limited to heavily parameterized cloud droplet sizes \citep{kitzmannetal2010,kopparapuetal2017}. Our results showcase that even slight changes in bulk cloud microphysical properties, such as cloud droplet size distributions, drastically change the location of the computed Inner Edge limits. For instance, an Earth-sized world's Inner Edge limit can change as much as 4 times the current insolation received by Earth, via changing only the sedimentation efficiency from 0.01 to 10.0 in the 100\% clouded cases. Further, the spread of changes in the the cloudy Inner Edge limits reported in \citet{selsisetal2007,yangetal2014,kopparapuetal2016,kopparapuetal2017}, can be accounted for by slight changes in the sedimentation efficiency (e.g., from $f_{\rm sed}$ = 0.32 to $f_{\rm sed}$ = 0.16). 

Our Inner Edge models also showcase a strong dependence on the planetary mass, similar to the results reported previously by \citet{kopparapuetal2014}. For instance, the \citet{kopparapuetal2014} results present a $\sim$ 7\% difference in effective incident starlight favoring super-Earth sized worlds, and our results indicate a comparable $\sim$ 8\% difference for low fractional cloudiness models (e.g., less than 75\%). Whereas, higher fractional cloudiness values (e.g., greater than 75\% ) the planetary mass dependence grows to as much as 20\%. This dependence on planetary mass is likely due to changes in atmospheric column depth as the atmospheric mass is scaled with planetary radius according to Equation (3) in \citet{kopparapuetal2014}.

The three parameters highlighted above: fractional cloudiness, sedimentation efficiency, and planetary mass, along with other planetary characteristics such as the global average surface temperature and atmospheric composition, contribute significantly to the climate characteristics of moist greenhouse worlds near the Inner Edge of the Habitable Zone. The myriad combinations of these parameters can lead to largely variable computed Inner Edge limits. For example, some of our highlighted results could imply that Early Venus was habitable in its early history \citep[e.g.,][]{kastingetal1993, Salvador2017}, in contrast to other reports suggesting that it was never habitable \citep{wolfeandtoon2014}. However, adopting similar model parameters to previous three-dimensional global climate models and intercomparing results is one key way to assess our moist greenhouse climate model's performance. For direct comparison, the \citet{yangetal2014} models suggest a global average surface temperature of 315~K for an atmospheric \ch{CO2} volume mixing ratio of 400 ppm, a surface gravity of 0.9$g_{\rm \oplus}$, and an insolation of $S_{\rm eff}$ = 2.64 for a Sun-like host. Adopting these same bulk planet parameter values, along with a mean Earth-like sedimentaion efficiency of $f_{\rm sed}$ = 0.32, and a fractional cloudiness of $f_{\rm cld}$ = 1.0 within our model results in a $S_{\rm eff}$ = 2.77, only a 5\% difference from the synchronous rotating, three-dimensional Inner Edge model of \citet{yangetal2014}.

\begin{figure}
    \centering
    \includegraphics[width=\textwidth,height=\textheight,keepaspectratio]{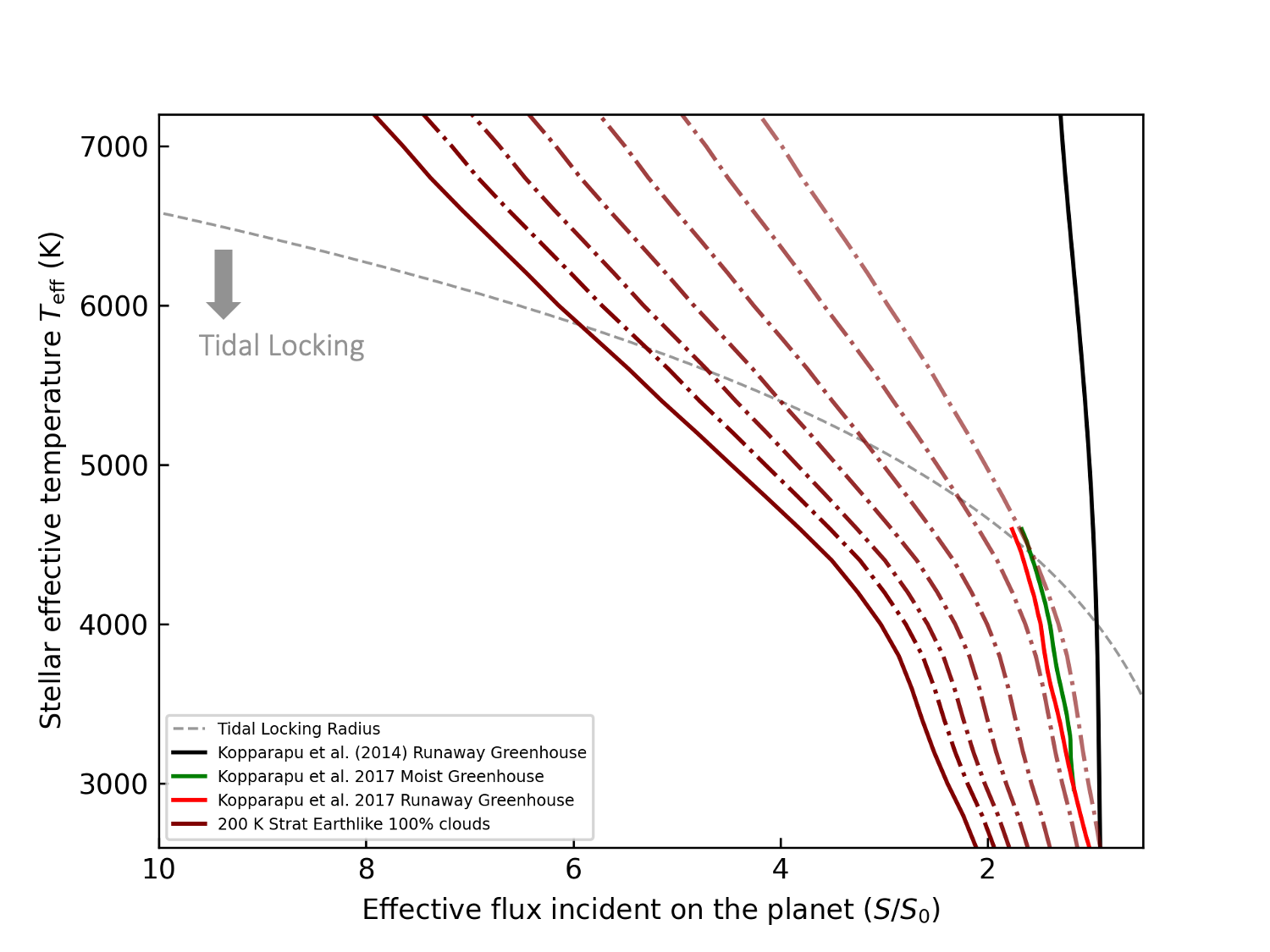}
    \caption{Inner Edge comparison to \citet{kopparapuetal2017}. Maroon lines represent 100\% clouded Earth-sized climates for sedimentation efficiency ranging from 0.01 to 0.64 in log spaced sequence from the leftmost solid line representing 0.01 and the rightmost dash-dot line representing a sedimentation efficiency of 0.64. The green solid line and red solid lines are from \citet{kopparapuetal2017} and are the moist greenhouse limit and Runaway Greenhouse limit, respectively. The two bordering (rightmost) inverse moist greenhouse climate states agree within 10\% of the inner edge results from \citet{kopparapuetal2017}. Here, they represent mean cloud droplet effective radii of 10 microns and 15 microns, whereas the \citet{kopparapuetal2017} clouds are tuned to Earth clouds and have effective cloud droplet radii around these values.}
    \label{fig:Cloud_rainout_seff} 
\end{figure}

\begin{table}
\centering
\begin{tabular}{ccc}
\hline
Constant                                     & 100\% Cloudy, $f_{\rm sed}$ = 0.32 & 100\% Cloudy, $f_{\rm sed}$ = 0.64 \\ \hline
$S_{\rm eff \odot}$ (1 $M_{\rm \oplus}$) & 3.3                                & 2.73                               \\
$S_{\rm eff \odot}$ (3.5 $M_{\rm \oplus}$) & 4.65                               & 4.09                               \\ \hline
a (1 $ M_{\rm \oplus}$)                      & 1.17 $\times 10^{\rm -3}$          & 1.01 $\times 10^{\rm -3}$          \\
a (3.5 $ M_{\rm \oplus}$)                      & 1.54 $\times 10^{\rm -3}$          & 1.37 $\times 10^{\rm -3}$          \\ \hline
b (1 $ M_{\rm \oplus}$)                      & 6.82 $\times 10^{\rm -8}$          & 7.94 $\times 10^{\rm -8}$          \\
b (3.5 $ M_{\rm \oplus}$)                      & 4.14 $\times 10^{\rm -8}$          & 5.31 $\times 10^{\rm -8}$          \\ \hline
c (1 $ M_{\rm \oplus}$)                      & -5.43 $\times 10^{\rm -11}$        & -4.16 $\times 10^{\rm -11}$        \\
c (3.5 $ M_{\rm \oplus}$)                      & -8.33 $\times 10^{\rm -11}$        & -6.90 $\times 10^{\rm -11}$        \\ \hline
d (1 $ M_{\rm \oplus}$)                      & -8.43 $\times 10^{\rm -15}$        & -6.90 $\times 10^{\rm -15}$        \\
d (3.5 $ M_{\rm \oplus}$)                      & -1.15 $\times 10^{\rm -14}$        & -9.74 $\times 10^{\rm -15}$       
\end{tabular}
\caption{Constants for the parametric fits to the Partially Clouded moist greenhouse Habitable Zone Limits. The values for Recent Venus, and Runaway Greenhouse are from \citet{kopparapuetal2014}, and are included for direct comparison. Here, only 100\% and 50\% clouded climates are fitted, as these represent a characteristic sample of cloud cover percentages that may arise on rocky worlds. We adopt Earth-like sedimentation efficiencies ($f_{\rm sed}$ = 0.32) and ($f_{\rm sed}$ = 0.64), for direct comparison to \citet{kopparapuetal2017}. }
\label{tab:Kopparapu_Comparison_Fit_coefficients}
\end{table}

For further model assessment, we compare the moist greenhouse Inner Edge limits of the Habitable Zone from \citet{kopparapuetal2017} to our grid of 100\% cloudy Earth-like inverse models spanning a sub-selection of sedimentation efficiency from 0.01 to 0.64 (Figure \ref{fig:Cloud_rainout_seff}). The most apparent result from directly comparing the two independent studies is the striking range in which these two studies predict the Inner Edge limits. For instance, our climate models for the most inefficient sedimentation efficiency place the inner edge for a late K dwarf with an effective temperature of 4000\,K at a stellar effective flux of 3.0\,$S_{\rm{eff}}$, whereas the \citet{kopparapuetal2017} Runaway Greenhouse limit is at a stellar effective flux of 1.5 $S_{\rm{eff}}$. However, the moist greenhouse Inner Edge inverse models more readily agree with the \citet{kopparapuetal2017} results if we adopt sedimentation efficiency values of 0.32 (left border) and 0.64 (right border), which correspond to mean effective cloud droplet radii of 10\,\textmu m and 15\,\textmu m, respectively. In \citet{kopparapuetal2017}, liquid cloud droplet radii are in all cases parameterized to be 15 \textmu m. Thus, for similar global average cloud cover and mean cloud droplet effective radii our cloudy one-dimensional results agree well with the three-dimensional global climate model results from \citet{kopparapuetal2017}, which uses parameterized uniform cloud droplet distributions of 15 \textmu m. 
It should be noted that the \citet{kopparapuetal2017} models neglect \ch{CO2} and other radiatively active trace gases for simplicity. In contrast, to keep the moist greenhouse Inner Edge limit calculations consistent with previous one-dimensional studies \citep{kastingetal1993,kopparapu2013,kopparapuetal2014}, we include a \ch{CO2} volume mixing ratio of 360 ppm. Here, discrepancies between mean cloud droplet effective radii of the three-dimensional global climate modeling Inner Edge limits in \citep{kopparapuetal2016,kopparapuetal2017} and our one-dimensional results, presented here, may be diminished by computing Inner Edge climates without \ch{CO2}. For example, for host stellar effective temperature of 4000~K, the disagreement between the Inner Edge limit decreases by 48\% when comparing the one-dimensional and three-dimensional models for the $f_{\rm sed}$ = 0.62 case if the \ch{CO2} volume mixing ratio is set to zero.

The mean water cloud droplet effective radii arising on worlds outside of our Solar System is not well constrained. Further, cloud cover parameters arising on rocky worlds are also poorly constrained, with many three-dimensional global climate models resulting in large variations in overall cloud cover despite controlled initial planetary parameters \citep{fouchezetal2021,sergeevetal2022}. Three-dimensional global climate models that are applied to surface ocean bearing moist greenhouse climate states specifically predict dense cloud decks around the substellar point for slow-rotators \citep{kopparapuetal2016,kopparapuetal2017} and \citet{kopparapuetal2017} also predict high fractional cloud cover for the entire planetary disk for a rapidly rotating moist greenhouse climate. Adopting a fractional cloudiness of 100\% in the slow-rotator regime captures the radiative environment of the slowly rotating worlds reported in \citet{kopparapuetal2017} -- here we are able to reproduce clouded Inner Edge results after accounting for mean effective cloud droplet radii. Despite no direct Inner Edge comparison to \citet{kopparapuetal2017} in the rapid-rotator regime, global cloud cover maps from \citet{kopparapuetal2017} showcase high cloud cover across the entire planetary disc, so we adopt a global cloud fraction of 100\% to approximate a similar cloud cover distribution. Adopting a similar sedimentation efficiency as for the slow-rotator regime of the moist greenhouse Inner Edge climate models results in a continuous Habitable Zone that can be extrapolated to stellar hosts with effective temperatures up to 7,200\,K. For a much more conservative estimate of the cloudy moist greenhouse Habitable Zone, we provide parametric fits to the \citet{kopparapuetal2017}-tuned inverse climate models  \ref{tab:Kopparapu_Comparison_Fit_coefficients}.



\subsection{Moist Greenhouse Planet Candidates}

Exoplanet observational capabilities are heavily dependant on planetary orbital distance from the host star \citep{kopparapuetal2017,madhusudhan2019}. In particular, the shorter orbital periods of Inner Edge planets make them prime targets for characterization studies, due to a higher potential cadence of follow-up observations and leading to additional data integration over a shorter period of time \citep{kopparapuetal2017}. Inner Edge worlds are likely to be warm and subject to potentially unstable climate states \citep{kastingetal1993,kopparapu2013,kopparapuetal2014,kopparapuetal2016,kopparapuetal2017}. A key driver of climate stability in the moist greenhouse regime is the radiative influence of cloud feedback mechanisms. Previously, we demonstrated that the presence of clouds can sextuple the amount of incident starlight a rocky planet orbiting a sunlike star could withstand and still remain habitable. Table~\ref{tab:Planet_Candidates} lists confirmed and unconfirmed rocky planet candidates that exist within the moist greenhouse regime outlined in this study. 

Atmospheric characterization of Habitable Zone worlds is a central goal of current and future rocky exoplanet observations. In adherence to this goal, large progress has been made with theoretical models of rocky exoplanet atmospheres which ultimately yield testable predictions of atmospheric composition, dynamics, and evolution. One of the main drivers of this significant progress is the possibility of characterizing worlds that harbor life through the presence of atmospheric biosignatures \citep{wordsworthetal2022}. Our study provides abiotic spectral models of rocky exoplanets near the Habitable Zone Inner Edge, where the self-consistent incorporation of condensate clouds makes the results valuable to retrieval modeling efforts. Honing the ability to characterize habitable abiotic exoplanets such as these\,---\,alongside worlds with spectral signatures driven by the presence of biology\,---\,is an important resource for interpreting real observations.


To understand the potential observability of moist greenhouse worlds, we look at recent JWST transit transmission spectroscopic observations of the potentially Venus-like rocky world, LHS-475b \citep{lustig-yaegeretal2023}. Here, we model the transit transmission spectrum of an Earth-sized cloudy moist greenhouse planet with a planetary radius of 1~R$_{\rm \oplus}$ and a 1~bar \ch{N2} background atmosphere. We assume a stellar spectral type of M3.5V with a stellar effective temperature of 3200\,K, and a stellar radius of 0.28\,R$_{\rm \odot}$. We diverge from true LHS-475b parameters and place the planet at an orbit of \,0.05~AU to simulate the moist greenhouse Inner Edge Limit for a 100\% cloudy Earth-sized world with a sedimentation efficiency of 0.64, specified by the parametric fits from Table~\ref{tab:Kopparapu_Comparison_Fit_coefficients}. The observations of LHS-475b reported by \citet{lustig-yaegeretal2023} occur over two transits and achieve a high enough signal to noise ratio to detect spectral features with a prominence of less than 50 ppm, over the spectral range of 2.9--5.3\,\textmu m and binned to 56 spectral points. As a theoretical exercise, we adopt a realistic noise floor of 50 ppm and construct a simple comparison to this real-world two-transit observation.


\begin{figure}
    \centering
    \includegraphics[width=\textwidth,height=\textheight,keepaspectratio]{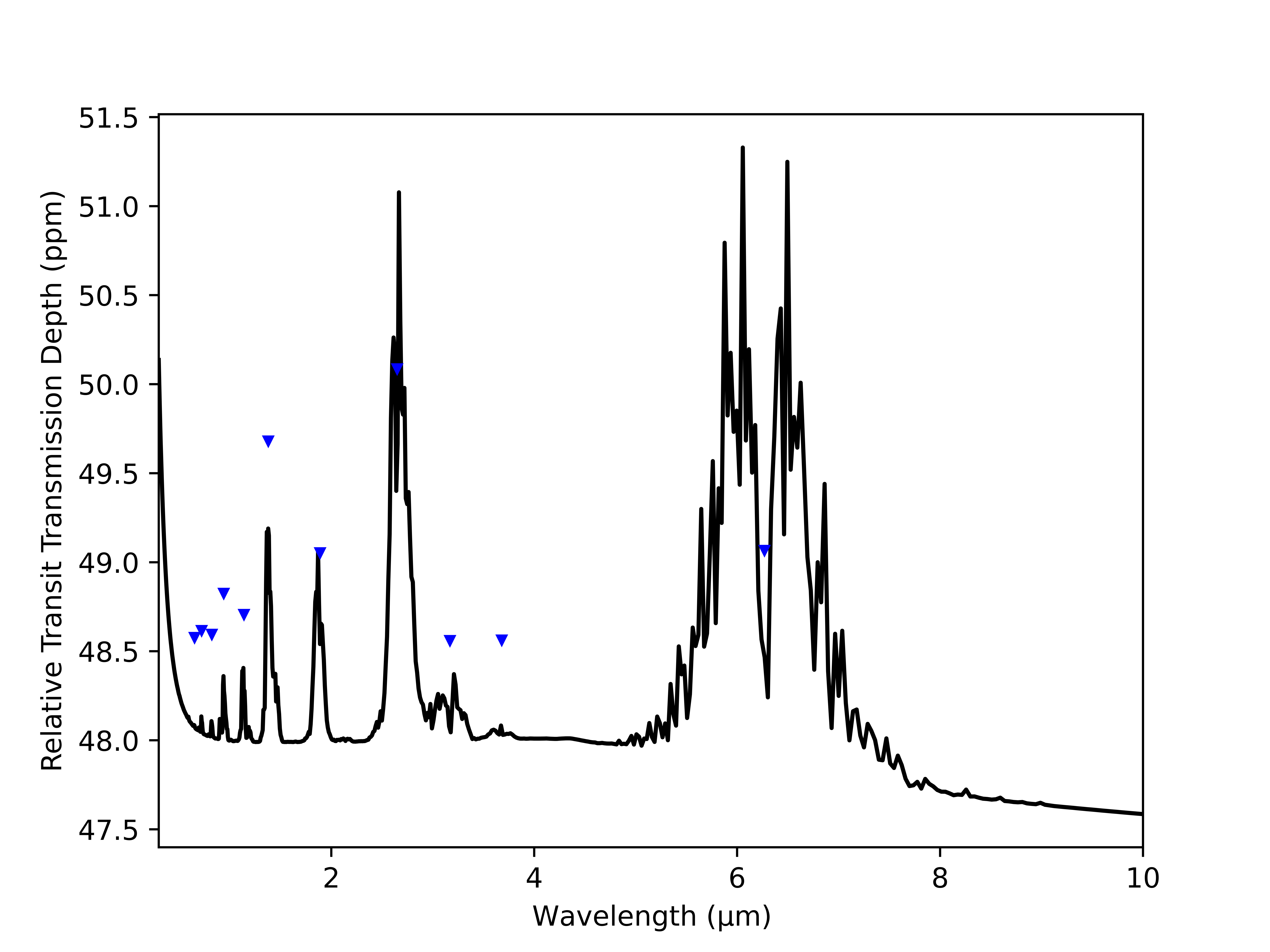}
    \caption{Normalized transit transmission spectrum from a coupled climate-to-spectral model of an Earth-sized moist greenhouse climate with 100\% cloud cover orbiting an M3.5V dwarf. Water vapor absorption band midpoints are labelled with  blue triangles. The most prominent features at 2.7 \textmu m and from 5--8 \textmu m do not surpass the 50 ppm observational uncertainty from the \citet{lustig-yaegeretal2019} data, with relative peaks above the continuum of only 2.5--3 ppm. }
    \label{fig:transit_inverse_expanded}
\end{figure}

\begin{figure}
    \centering
    \includegraphics{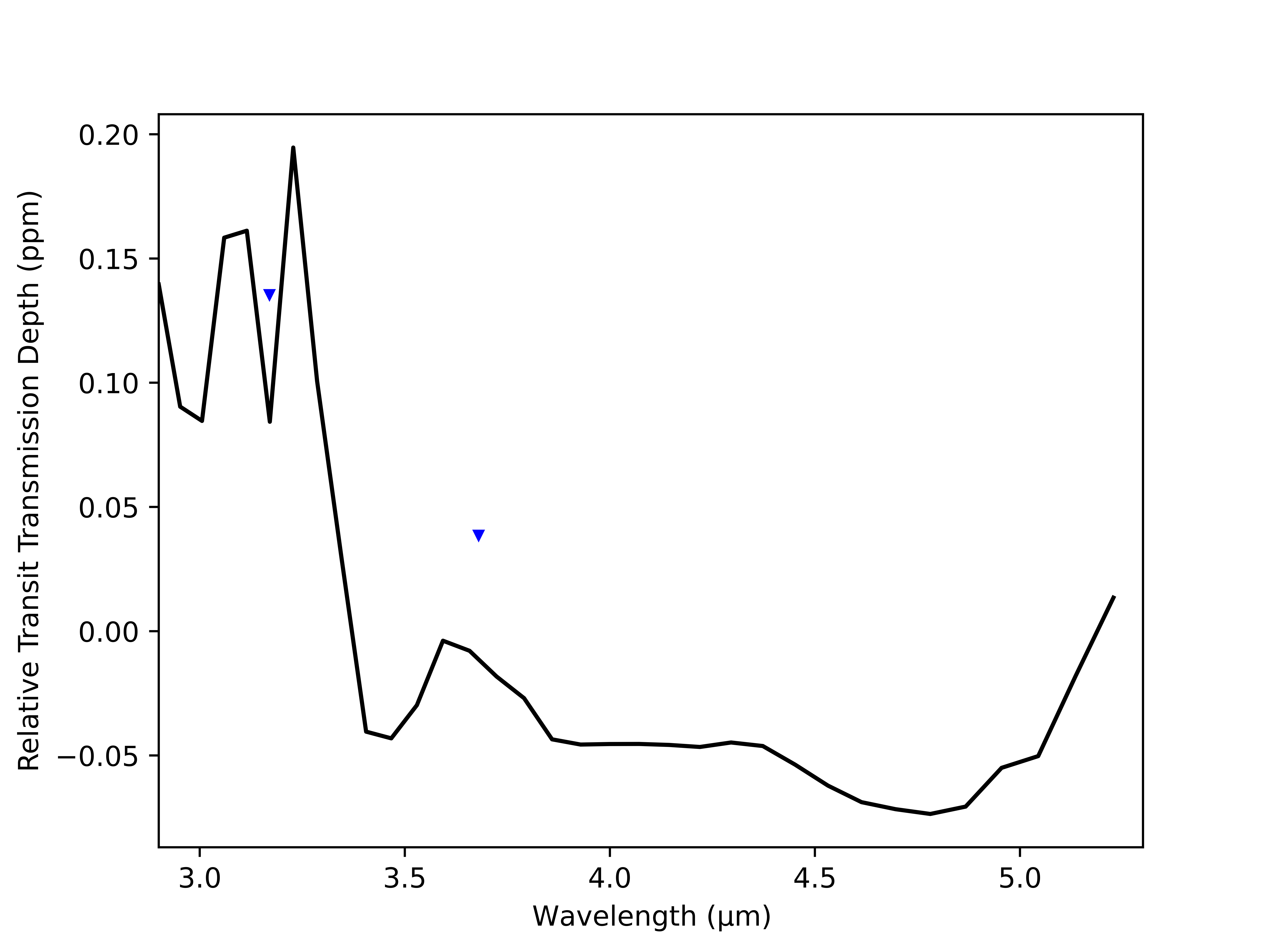}
    \caption{Binned normalized transit transmission spectrum from the Earth-sized moist greenhouse climate with 100\% cloud cover orbiting an M3.5V dwarf in Figure \ref{fig:transit_inverse_expanded}. Here the spectral model is binned to 56 data points from 2.9 to 5.3 \textmu m and is a representation of a JWST two-transit integrated observation of an rocky Earth-sized world in a moist greenhouse climate state. }
    \label{fig:transit_inverse_relative}
\end{figure}

 Figure~\ref{fig:transit_inverse_expanded} showcases the normalized relative transit depth, where all spectral features are normalized by the mean transit depth. Transit transmission spectrum models of moist greenhouse worlds without \ch{CO2} absorption and 100\% cloud cover are monopolized by \ch{H2O} spectral features above the cloud deck, namely spectral absorption from 5--8\,\textmu m and at 2.7\,\textmu m. Here, the strongest spectral features result in relative transit depths of less than 10s of ppm and would be far below the 50 ppm observational uncertainty from the \citet{lustig-yaegeretal2019} data. Mimicking the observations in \citet{lustig-yaegeretal2023}, we bin the spectral models to 56 datapoints between 2.9 and 5.3 \textmu m in Figure~\ref{fig:transit_inverse_relative}. The resulting relative transit transmission spectrum is relatively flat, with few discernible molecular absorption features above the cloud deck for the majority of the wavelength range. This exercise suggests that observing a cloudy moist greenhouse climate state with only two transits with the JWST is not likely to lead to (full) atmospheric characterization. Further, as apparent in Figure~\ref{fig:reflection}, direct imaging studies may be more useful in characterizing cloudy moist greenhouse climate states.

\begin{table}
\resizebox{\textwidth}{!}{%
\begin{tabular}{lllccccc}
\hline
Planet Name             & Status                                        & Planetary Parameter Reference                                                                                                & Planet Radius (R$ _{\oplus}$) & Insolation Flux ($S_{\rm {eff}}$) & Stellar Effective Temperature (K) & Stellar Radius (R$_{\rm \odot}$)  & Distance (pc) \\ \hline
Kepler-1633 b           & Kepler Project Candidate (q1\_q17\_dr24\_koi) & \href{https://exoplanetarchive.ipac.caltech.edu/docs/Kepler\_KOI\_docs.html}{exoplanetarchive.ipac.caltech.edu}              & 1.06                          & 3.01                              & 6351.0                            & 0.92                              & 1114.91       \\
Kepler-1126 c           & Kepler Project Candidate (q1\_q17\_dr24\_koi) & \href{https://exoplanetarchive.ipac.caltech.edu/docs/Kepler\_KOI\_docs.html}{exoplanetarchive.ipac.caltech.edu}              & 1.54                          & 2.07                              & 6106.0                            & 0.84                              & 635.736       \\
Kepler-1126 b           & Kepler Project Candidate (q1\_q16\_koi)       & \href{https://exoplanetarchive.ipac.caltech.edu/docs/Kepler\_KOI\_docs.html}{exoplanetarchive.ipac.caltech.edu}              & 1.58                          & 4.67                              & 6106.0                            & 0.84                              & 635.736       \\
Kepler-442 b            & Kepler Project Candidate (q1\_q16\_koi)       & \href{https://exoplanetarchive.ipac.caltech.edu/docs/Kepler\_KOI\_docs.html}{exoplanetarchive.ipac.caltech.edu}              & 1.56                          & 1.08                              & 4569.0                            & 0.65                              & 365.965       \\
Kepler-437 b            & Kepler Project Candidate (q1\_q12\_koi)       & \href{https://exoplanetarchive.ipac.caltech.edu/docs/Kepler\_KOI\_docs.html}{exoplanetarchive.ipac.caltech.edu}              & 1.37                          & 1.78                              & 4427.0                            & 0.62                              & 417.0 $\pm$ 24.0\\
Kepler-1450 b           & Kepler Project Candidate (q1\_q16\_koi)       & \href{https://exoplanetarchive.ipac.caltech.edu/docs/Kepler\_KOI\_docs.html}{exoplanetarchive.ipac.caltech.edu}              & 1.55                          & 2.19                              & 4405.0                            & 0.61                              & 505.829       \\
Kepler-395 c            & Published Confirmed                           & \citet{torresetal2017}                                                                                                       & 1.336                         & 2.11                              & 3931.0                            & 0.55                              & 421.382       \\
K2-3 d                  & Published Confirmed                           & \citet{crossfieldetal2015}                                                                                                   & 1.52                          & 1.51                              & 3896.0                            & 0.56                              & 44.0727       \\
K2-3 c                  & Published Confirmed                           & \citet{crossfieldetal2016}                                                                                                   & 1.18                          & 1.77                              & 3841.0                            & 0.37                              & 44.0727       \\
Kepler-395 c            & Kepler Project Candidate (q1\_q17\_dr25\_koi) & \href{https://exoplanetarchive.ipac.caltech.edu/docs/Kepler\_KOI\_docs.html}{exoplanetarchive.ipac.caltech.edu}              & 1.14                          & 1.71                              & 3765.0                            & 0.52                              & 421.382       \\
Kepler-438 b            & Published Confirmed                           & \citet{dressingetal2015}                                                                                                     & 1.01                          & 1.31                              & 3747.0                            & 0.66                              & 196.0 $\pm$ 5.0 \\
Kepler-296 e            & Published Confirmed                           & \citet{barclayetal2015}                                                                                                      & 1.53                          & 1.41                              & 3740.0                            & 0.48                              & 558.0         \\
Kepler-296 d            & Kepler Project Candidate (q1\_q17\_dr25\_koi) & \href{https://exoplanetarchive.ipac.caltech.edu/docs/Kepler\_KOI\_docs.html}{exoplanetarchive.ipac.caltech.edu}              & 1.52                          & 1.83                              & 3740.0                            & 0.48                              & 558.0         \\
Kepler-395 c            & Published Confirmed                           & \citet{dressingetal2013}                                                                                                     & 1.18                          & 1.15                              & 3735.0                            & 0.4                               & 421.382       \\
Kepler-54 d             & Published Confirmed                           & \citet{dressingetal2013}                                                                                                     & 1.14                          & 1.47                              & 3579.0                            & 0.33                              & 271.701       \\
K2-72 c                 & Published Confirmed                           & \citet{crossfieldetal2016}                                                                                                   & 0.86                          & 1.41                              & 3497.0                            & 0.23                              & 66.4321       \\
Kepler-1512 b           & Published Confirmed                           & \citet{torresetal2017}                                                                                                       & 0.8                           & 1.6                               & 3484.0                            & 0.39                              & 315.0 $\pm$ 24.0\\
TOI-700 e               & Published Confirmed                           & \citet{gilbertetal2023}                                                                                                      & 0.953                         & 1.27                              & 3459.0                            & 0.42                              & 31.1265       \\
Kepler-560 b            & Published Confirmed                           & \citet{torresetal2017}                                                                                                       & 1.47                          & 1.18                              & 3430.0                            & 0.27                              & 109.308       \\
K2-72 e                 & Published Confirmed                           & \citet{dressingetal2017}                                                                                                     & 1.29                          & 1.2                               & 3360.47                           & 0.33                              & 66.4321       \\
TRAPPIST-1 d            & Published Confirmed                           & \citet{agoletal2021}                                                                                                         & 0.788                         & 1.11                              & 2566.0                            & 0.12                              & 12.43 $\pm$ 0.02             

\end{tabular}%
}
\caption{Moist greenhouse planet candidates. A catalogue of confirmed and unconfirmed rocky planets with orbital distances within the cloudy moist greenhouse limit reported in this work. The table is organized according to stellar effective temperature decreasing downward.}
\label{tab:Planet_Candidates}
\end{table}

\section{Conclusions}\label{sec:Conclusion}

Clouds fundamentally influence the location of the Inner Edge of the Habitable Zone. With a new climate model developed and validated in \citet{windsoretal2022}, we explored the role of convective clouds in moist greenhouse climate states and computed the cloudy Inner Edge(s) of the Habitable Zone for main sequence stars following classical Habitable Zone studies methods \citep{kastingetal1993,kopparapuetal2013,kopparapuetal2014}. Our major findings are:

\begin{enumerate}

    \item For the first time, we compute cloudy Inner Edge of the Habitable Zone limits with a parameterized microphysics cloud model coupled to a 1D inverse climate model. We find that the cloudy Inner Edge limits of the Habitable Zone are primarily dependent on four main model parameters: instellation, global fractional cloudiness ($f_{\rm cloud}$), sedimentation efficiency ($f_{\rm sed}$), and planet radius ($R_{\rm p}$). Changes in global fractional cloudiness can move the Inner Edge of the Habitable Zone for Earth-sized planets orbiting Sun-like stars from just inside the Earth's orbit (0.97 AU, $S_{\rm eff}$ = 1.06) at the lowest end, to between the orbit of Mercury and Venus (0.42 AU, $S_{\rm eff}$ = 5.8), at the high end. Increasing the cloud sedimentation efficiency to account for extremely efficient sedimentation can move this Inner Edge limit to 0.75 AU ($S_{\rm eff}$ = 1.78) (in an extreme case of $f_{\rm sed}$ = 10.0) via greatly increasing cloud droplet sizes, and decreasing cloud altitude, optical depth, and scattering efficiency in the visible. 
    
    \item Additionally, we note a dependence on total atmospheric mass for the Inner Edge limits, which on average accounts for a $\sim$ 8\% difference in $S_{eff}$, well in agreement with findings from \citet{kopparapuetal2014}. However, for high fractional cloudiness, as is expected for moist greenhouse climates \citep{yangetal2014,kopparapuetal2016,kopparapuetal2017}, the dependence on total atmospheric mass as a function of planetary mass grows to $\sim$ 20\%, driving 100\% clouded super-Earth Inner Edge limits 10\% closer to their stellar hosts.

    \item We compare inverse climate modeling moist greenhouse Inner Edge limit(s) directly to sophisticated three-dimensional global climate models of moist greenhouse Inner Edge worlds from \citet{kopparapuetal2017}. Here, we find that our inverse climate model predicts Inner Edge limits within 10\% of previous three-dimensional global climate model studies when tuning bulk cloud properties to be complementary. 

    \item A number of rocky planet candidates meet the requirements outlined in this study to potentially be Inner Edge moist greenhouse planets. Observations of these worlds would be valuable for constraining models of planetary habitability. Transit transmission spectroscopic modeling of clouded moist greenhouse climate states reveal that \ch{CO2} should be one of the more prominent spectral features above the continuum from the cloud deck, even for atmospheres where \ch{CO2} is a trace component. The \ch{CO2} absorption amplitudes above the continuum can provide information on physical cloud properties, including relative cloud deck altitude and fractional cloudiness. Additional statistical analysis of this relationship is required to further quantify these promising spectral characteristics.

\end{enumerate}

\section*{Acknowledgments}
TDR and JDW gratefully acknowledge support from NASA's Exoplanets Research Program (No.~80NSSC18K0349). TDR also acknowledges support from NASA's Exobiology Program (No.~80NSSC19K0473) and the Cottrell Scholar Program administered by the Research Corporation for Science Advancement. AS and TDR acknowledge support from NASA's Habitable Worlds Program (No.~80NSSC20K0226). TDR and VSM acknowledge support from the NExSS Virtual Planetary Laboratory Team, funded by NASA Astrobiology Program Grant No.~80NSSC18K0829. RK and AVY acknowledge support from the GSFC Sellers Exoplanet Environments Collaboration (SEEC), which is supported by NASA's Planetary Science Division's Research Program. The authors thank Schuyler Borges for thoughtful advice on figure and material presentation.


%

%

\vspace{5mm}



\bibliographystyle{aasjournal}
\bibliography{sample631}

\end{document}